\documentclass[fleqn,usenatbib]{mnras}

\usepackage{newtxtext,newtxmath}

\usepackage[T1]{fontenc}

\DeclareRobustCommand{\VAN}[3]{#2}
\let\VANthebibliography\thebibliography
\def\thebibliography{\DeclareRobustCommand{\VAN}[3]{##3}\VANthebibliography}

\usepackage{amsmath}
\usepackage{graphicx}
\usepackage{float}

\usepackage{amssymb}	
\usepackage{makecell}
\usepackage{orcidlink}
\usepackage{soul}
\usepackage{anyfontsize}

\newcommand{\smpc}{\,h^{-1}\mathrm{Mpc}}
\newcommand{\kmpc}{\,\mathrm{Mpc}^{-1}h}

\newcommand{\ourModel}{network model }
\newcommand{\fog}{FoG }
\newcommand{\rec}{reconstructed } 

\usepackage{xcolor}
\usepackage{ulem,times}
\newcommand\mysout{\bgroup\markoverwith{\textcolor{black}{\rule[0.5ex]{2pt}{0.6pt}}}\ULon}

\definecolor{ForestGreen}{rgb}{0.3,0.7,0.3}

\title[Peculiar velocity and RSD]{
Estimation of line-of-sight velocities of individual galaxies using neural networks I. Modelling redshift-space distortions at large scales}

\author[Hongxiang Chen et al.]{
Hongxiang Chen$^{1,2}$\thanks{E-mail: chxiang@nao.cas.cn}~\orcidlink{0009-0001-2951-7383},
Jie Wang$^{1,2}$\thanks{E-mail:  jie.wang@nao.cas.cn},
Tianxiang Mao$^{1,2}$~\orcidlink{0000-0001-6772-9814},
Juntao Ma$^{1,2}$~\orcidlink{0009-0000-5513-4100},
Yuxi Meng$^{1,2}$~\orcidlink{0009-0004-6895-6743},
Baojiu Li$^{3}$~\orcidlink{0000-0002-1098-9188},
\newauthor{
Yan-Chuan Cai$^{4}$,
Mark Neyrinck$^{5,6}$~\orcidlink{0000-0002-2618-5790},
Bridget Falck$^{7}$,
Alexander S.\ Szalay$^{8}$}
\\
$^{1}$National Astronomical Observatories, Chinese Academy of Sciences, Beĳing, 100101, China\\
$^{2}$School of Astronomy and Space Science, University of Chinese Academy of Sciences, Beĳing 100049, China\\
$^{3}$Institute for Computational Cosmology, Department of Physics, Durham University, Durham DH1 3LE, UK\\
$^{4}$Institute for Astronomy, University of Edinburgh, Royal Observatory, Blackford Hill, Edinburgh EH9 3HJ, UK\\
$^{5}$Department of Physics and Astronomy, University of Denver, Denver, CO, 80210, USA\\
$^{6}$Blue Marble Space Institute of Science, Seattle, WA, 98104, USA\\
$^{7}$Independent Researcher, Baltimore, MD 21201, USA\\
$^{8}$Department of Physics and Astronomy, The Johns Hopkins University, Baltimore, MD 21218, USA\\
}

\date{Accepted XXX. Received YYY; in original form ZZZ}

\pubyear{\the\year{}}

\begin{document}
\label{firstpage}
\pagerange{\pageref{firstpage}--\pageref{lastpage}}
\maketitle

\begin{abstract}
 We present a scheme based on artificial neural networks (ANN) to estimate the line-of-sight velocities of individual galaxies from an observed redshift-space galaxy distribution. 
 {We find an estimate of the peculiar velocity at a galaxy based on galaxy
counts and barycenters in shells around it. By training the network with environmental characteristics, such as the total mass and mass center within each shell surrounding every galaxy in redshift space,} our ANN model can accurately predict the line-of-sight velocity of each individual galaxy. When this velocity is used to eliminate the RSD effect, the two-point correlation function (TPCF) in real space can be recovered with an accuracy better than 1\% at $s$ > 8 $\smpc$, and 4\% on all scales compared to ground truth. The real-space power spectrum can be recovered within 3\% on $k$< 0.5 $\kmpc$, and less than 5\% for all $k$ modes. The quadrupole moment of the TPCF or power spectrum is almost zero down to $s$ = 10 $\smpc$ or all $k$ modes , indicating an effective correction of the spatial anisotropy caused by the RSD effect. We demonstrate that on large scales, without additional training with new data, our network is adaptable to {different galaxy formation models}, different cosmological models, and mock galaxy samples at high redshifts and high biases, achieving less than 10\% error for scales greater than 15 $\smpc$. As it is sensitive to large-scale densities, it does not manage to remove Fingers of God in large clusters, but works remarkably well at recovering real-space galaxy positions elsewhere. Our scheme provides a novel way to predict the peculiar velocity of individual galaxies, to eliminate the RSD effect directly in future large galaxy surveys, and to reconstruct the 3-D cosmic velocity field accurately.

\end{abstract}




\section{Introduction}

Galaxy positions and velocities contain rich and important information, to constrain cosmological models. For instance, estimation of the peculiar velocity field of galaxies or bulk flows can provide effective constraints \citep{Strauss1995,Erdogdu2006,Ma2012, Springob2016, Carrick2015, Boruah2020, Lilow2021}. However, the line-of-sight velocity is coupled with the line-of-sight distance (redshift) for galaxies in observations, especially for faraway objects. This phenomenon is commonly referred to as the redshift-space distortion effect (hereafter RSD effect) and must be taken into account for cosmological analyses of galaxy position data, such as in analysis of Baryon Acoustic Oscillations (BAO).

Normally, the peculiar velocity of a galaxy can be derived from the matter distribution around. In a rudimentary sense, the peculiar velocity of a galaxy can be divided into two parts: the peculiar velocity of the galaxy's host dark matter halo, determined by the large-scale matter distribution, and the random velocity of the galaxy within the dark halo.
The RSD effect can approximately be decomposed into
the Kaiser effect \citep{Kaiser1987,Hamilton1998,Szalay1998} and the {F}inger-of-God effect (FoG) \citep{Jackson1972,Kaiser1987}.

On large scales, it is well established that there exists a robust correlation between the velocity distribution and the matter density of the universe \citep{Peacock1994, Scoccimarro2004, Taruya2010, Seljak2011, Okumura2012a, Okumura2012b}. Therefore, the cosmic velocity field can be inferred from the matter field
\citep{Croft1997, Nusser1991, Branchini1999, Nusser1994,Fisher1995, Branchini2002, Landy2002, Erdogdu2006, kudlicki2000, Kitaura2012,Zhang2013, Tanimura2022, GaneshaiahVeena2022, Wu2023, Qin2023}. Nevertheless, the results of these works are generally limited by the fact that the linear relationship between the velocity field and the mass distribution is weak. An iterative velocity-field reconstruction approach has had some success \citep{Yahil1991, Wang2009, Kitaura2012, Wang2012, Shi2016, Yu2019, Wang2020}.

Most of these existing schemes provide measurements or reconstruction of the cosmic velocity field on a grid. The velocity of individual galaxies is then interpolated from the grid to the galaxies' positions. An accurate estimation of the peculiar velocity of an individual galaxy requires a detailed description of the matter distribution around this galaxy. Normally, the relation between an individual galaxy and its environment is highly nonlinear, and the estimation of the peculiar velocity for each and every observed galaxy will make the computation complex and expensive. In light of these challenges, artificial neural networks (ANN) are an efficient nonlinear and data-driven algorithm \citep{goodfellow2016deep} to deal with this kind of problem.  ANNs have been introduced into the field of cosmology and have achieved many successes in improving accuracy and precision in constraining cosmological models \citep{Ravanbakhsh2017, Schmelzle2017,Fluri2018,Peel2019PhysRevD, Merten2019, Mao2021, Villaescusa-Navarro2022,de-Santi2023,Shao2023, Huertas2023}. For example, using ANN, we have proposed a new scheme to reconstruct the BAO signal, resulting in a significant improvement in the signal-to-noise ratio of BAO down to small scales \citep{Mao2021}. 

In light of these successes, it is worthwhile to try to estimate the line-of-sight velocities of individual galaxies using an ANN. Given that the RSD effects at large and small scales are determined by the large-scale density and virial motion in the halo, respectively, we can try to model the RSD at these two different scales separately. In this paper, we focus on modelling the RSD effect at large scales, especially on the line-of-sight velocity. In a parallel paper, we will show a graph neural network-based \fog model that focusses on extracting the small scale information.

This paper is organised as follows. In Section \ref{sec:method} we describe our peculiar-velocity model based on the neural network and the simulations used in this work. Then we discuss the results of the velocity prediction and RSD correction in Section \ref{sec:results}. We discuss the cosmology dependence in Section \ref{sec:discussion}, and finally conclude in Section \ref{sec:conclusion}. As a series of works, in a subsequent article, we will focus on how to eliminate the \fog effect caused by virial motions in galaxy clusters.

\section{METHOD} \label{sec:method}

In this section, we introduce our new scheme to estimate the line-of-sight velocity of galaxies using artificial neural networks (ANN) and the modelling of the RSD effect. 

\subsection{Peculiar-velocity and Redshift-space distortion}

In observations, the distance between a galaxy and us is measured by its redshift $z_{\rm obs}$. $z_{\rm obs}$ includes not only the cosmological redshift $z_{\rm cos}$, which arises from the Hubble expansion and implies the real galaxy position, but also the Doppler redshift $z_{\rm pec}$, which is caused by the peculiar velocity of the galaxy itself along the line-of-sight $v_{\mathrm{los}}$,

\begin{equation}
\begin{split}
    &\lambda_{\rm obs}/\lambda_{\rm em}=1+z_{\rm obs}=(1+z_{\rm doppler})(1+z_{\rm cos}) \approx 1+z_{\rm cos}+z_{\rm pec}, \\
    &\rm{if} \quad z_{\rm cos}, z_{\rm pec}\ll1\\
\end{split}
\end{equation}

The peculiar velocity of a galaxy significantly distorts its position in
observations, and such an RSD effect will bring about the anisotropy signal in
the observed distribution of galaxies. The position of a galaxy inferred by $z_{\rm obs}$, usually referred to as its position in redshift space, is related to its
real position in the universe as 
\begin{equation}\boldsymbol{s}=\boldsymbol{x}+\frac{v_{\mathrm{los}}}{a H(a)} \vec n,
    \label{eq: formulartwo}
\end{equation}
where $\boldsymbol{s}$ and $\boldsymbol{x}$ are the coordinates in redshift and real space
respectively, $\vec n$ is the unit vector for the line-of-sight direction, $v_{\mathrm{los}}$
identifies the component of peculiar velocity in the line of sight, $a$ is the scale factor
and $H(a)$ is the Hubble expansion rate at $a$.

The anisotropy of galaxy distribution caused by redshift-space distortions can be found on both large and small scales. On large scales,  matter around the overdensity region is falling in, and such a stream velocity will make the scatter of $z_{\rm obs}$ for those galaxies in this region smaller than that of their $z_{\rm cos}$, resulting in a visually compression effect along
the line-of-sight, known as the Kaiser effect \citep{Kaiser1987}. On small scales,
some nearby galaxies around the target galaxy at the region centre, can have very large line-of-sight peculiar velocities due to virial motion, and will
make the scatter of $z_{\rm obs}$ for those galaxies along the line-of-sight much bigger
than that of their $z_{\rm cos}$,  which makes these galaxies stretch seriously along
the line-of-sight,like the finger of God, known as the \fog effect
\citep{Jackson1972,Kaiser1987}.

A galaxy's peculiar velocity is determined by the matter density around it, which differs at different times as the Universe evolves. In the linear approximation, the relationship between the velocity and density fields on large scales is given by
\begin{equation}
    \boldsymbol{v}(\boldsymbol{k})=H a f(a) \frac{i \boldsymbol{k}}{k^{2}} \delta(\boldsymbol{k}).
    \label{eq: formularthree}
\end{equation}
Here, the $\boldsymbol{v}(\boldsymbol{k})$ and $\delta(\boldsymbol{k})$ are the velocity and
density fields in Fourier space, respectively, and $f(a)$ the linear growth rate defined as
$f(a)=d \ln D / d \ln a \simeq \Omega_{\mathrm{m}}^{0.6}+\frac{1}{70} \Omega_{\Lambda}\left(1+\Omega_{\mathrm{m}} / 2\right)$ \citep[e.g.][]{lahav1991}.

Correspondingly, we can divide a galaxy's peculiar velocity into two parts. One part is caused by the large scale structure in the universe, which is the source of the large-scale RSD. Another part is determined by the local environment, especially the dark matter halo where the galaxies are embedded. In this paper, we focus only on the modelling of RSD at large scales, especially on the line-of-sight velocity; the modelling at small scales will be presented in a parallel paper later. 

\subsection{A naive model of peculiar velocity}

Modelling a galaxy's peculiar velocity accurately is always a great challenge. Traditional methods, such as perturbation theory and iterative RSD correction, can correct the large scale linear RSD and extend it to weak linear scales, but they generally work less well at small, strongly, nonlinear scales. In order to work better on the latter scales, in this work we introduce a new method -- artificial neural networks. 

The first question we need to answer for such a method is a physics-orientated one: What information do we need to feed into the networks for training. We try to set up a naive model below to see what information is essential to model peculiar velocities of galaxies physically. 



To extract the line-of-sight velocity of galaxies from survey data, we need to define variables that contain the necessary information. It is believed that a galaxiy's peculiar velocity is determined by surrounding matter, which is normally traced by neighbouring galaxies. After several attempts, we design a spherical shell model to describe the environment of a galaxy. In the following, we introduce the details of the spherical shell model.

For each galaxy (hereafter target galaxy), we consider the gravitational effect of neighbouring galaxies at different radial distances. We divide these galaxies into different subpopulations in different radius bins. Then, we calculate the acceleration contributed by each spherical shell to the target galaxy. This set of accelerations describes the environment of the target galaxy at different scales. We write the relationship between the velocity and accelerations of the target galaxy as
\begin{equation}
    \boldsymbol{v}=f\left(\boldsymbol{a_{1}}, \boldsymbol{a_{2}}, \ldots, \boldsymbol{a_{n}}\right),
\end{equation}
where
\begin{equation}
    \boldsymbol{a_{j}}=\sum_{r_{j}<r_{k}<r_{j+1}}^{k} \frac{G m_{k} \vec{\boldsymbol{x}}_{k}}{r_{k}^{3}}.
\end{equation}
Among them, the function $f$ is an unknown nonlinear function, and we will use a \ourModel to approximate it; \boldmath {$a_{j}$} \unboldmath is the gravitational acceleration contributed by all galaxies in the $j$-th spherical shell to the target galaxy; the subscript $k$ refers to all galaxies in the shell $r_j$ to $r_{j+1}$, and $n$ is the total number of shells. For simplicity of calculation, we make some simplifications. When the shell is thin enough, $r_k$ can be replaced by $r_j$:
\begin{equation}
    \boldsymbol{a}_{j}=\sum_{r_{j}<r_{k}<r_{j+1}}^{k} \frac{G m_{k} \boldsymbol{x}_{k}}{r_{k}^{3}} \approx \frac{G}{r_{j}^{3}} \sum_{r_{j}<r_{k}<r_{j+1}}^{k} m_{k} \boldsymbol{x}_{k} \approx \frac{G}{r_{j}^{3}} M_{j} \cdot \boldsymbol{X}_{j},
    \label{eq: formularsix}
\end{equation}
where $M_j$ and $X_j$ are the mass and barycenter of the $j$-th spherical shell, respectively. As a result, for each target galaxy, the gravitational accelerations of different scales are proportional to the product of the shell mass and the mass centre of the shell.

Therefore, a galaxy's peculiar velocity can be written as a function of the mass and barycenter of each spherical shell
\begin{equation}
    \boldsymbol {v}=f\left(M_{1}, \boldsymbol {X}_{1}, M_{2}, \boldsymbol {X}_{2}, \ldots, M_{n}, \boldsymbol {X}_{n}\right)
    \label{eq: formulareight}.
\end{equation}

However, the real mass of each galaxy is difficult to predict accurately. This problem introduces some errors into our model that are difficult to quantify. Therefore, considering the divergences between different galaxy formation models and bearing in mind future application of our method in observations, it is worth noting that when calculating the input spherical shell model of neural network, the mass and barycenter of each spherical shell, are calculated by using the count of subhalos in simulation, instead of being weighted by real subhalo mass. In the same way, we assume that each subhalo above a certain mass limit will produce a galaxy, and in this work the mass cut is $2.58 \times 10^{12} h^{-1} \mathrm{M}_{\odot}$. So $M_i$  can be replaced by the count of subhalo $N_i$ and the barycenter $X_i$ is also calculated by the count of subhalos. So Eq.~(\ref{eq: formulareight}) can be written as:
\begin{equation}
    \boldsymbol {v}=f\left(N_{1}, \boldsymbol {X}_{1}, N_{2}, \boldsymbol {X}_{2}, \ldots, N_{n}, \boldsymbol {X}_{n}\right).
    \label{eq: formularnine}
\end{equation}

\begin{figure*}
	\includegraphics[width=\textwidth]{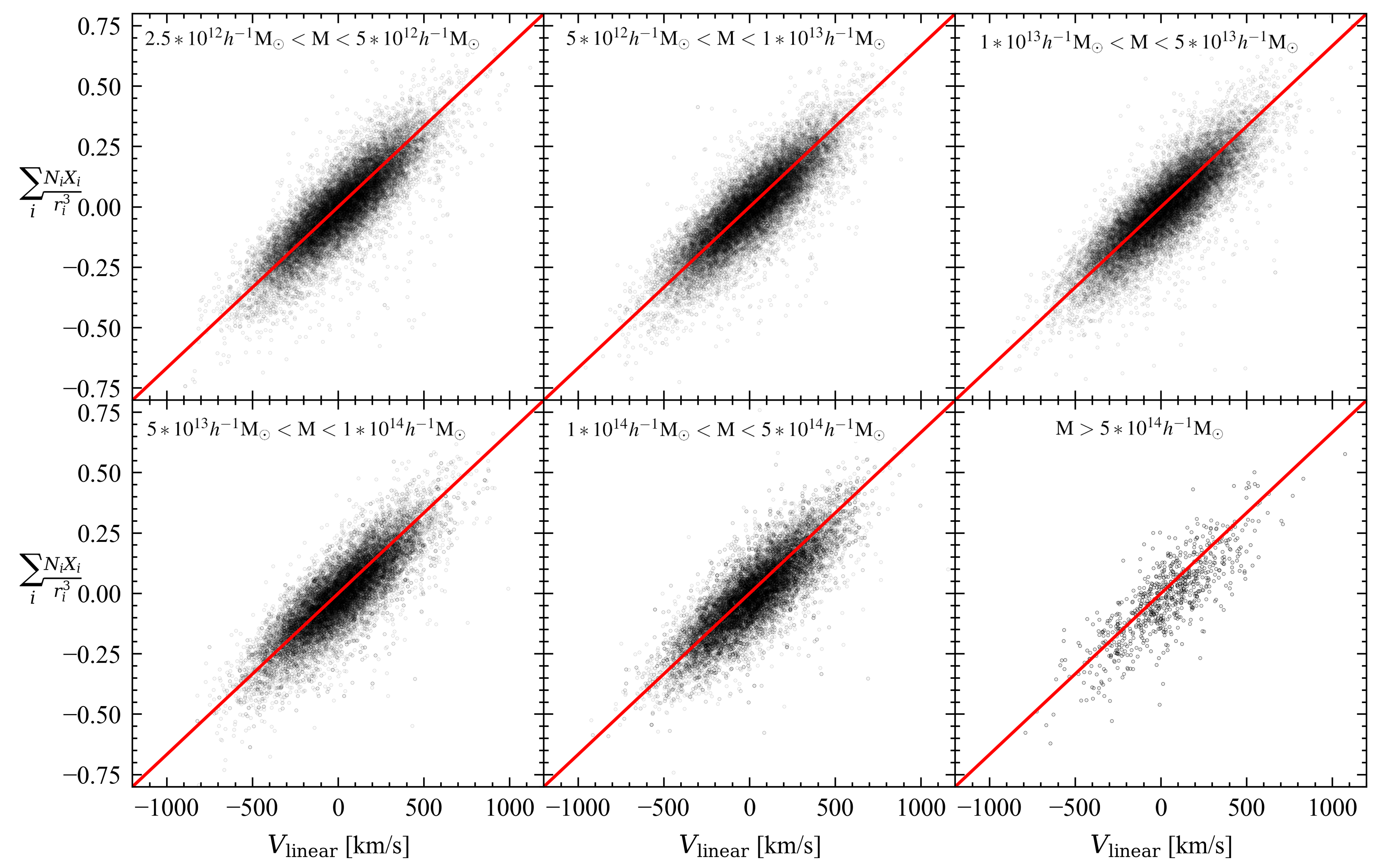}
    \caption{{The comparison between the naive model outcomes and the linear velocities for each galaxy. Across all mass ranges, the model outcomes align closely with the linear velocities with an acceptable scatter. This suggests that the linear combination of galaxy counts and barycenters within the spherical shells effectively predicts galaxy velocities.}}
    \label{fig:figure0_theory}
\end{figure*}

{ Combining Eq.~(\ref{eq: formularsix}) and Eq.~(\ref{eq: formularnine}), we can define the acceleration in our naive model as the linear combination of counts and barycenters of the spherical shells: 
\begin{equation}
    \sum_{i} {a}_{i}=\sum_i \frac{N_i X_i}{r_i^3}.
    \label{eq: formularten}
\end{equation}
The calculation of $\sum_i \frac{N_i X_i}{r_i^3}$ used all spherical shells bigger than 6.4 $\smpc$ in Table~\ref{tab:table1}.}

{It will be interesting to check the performance of the naive model. To do this, we can compare the outcomes of our naive model (Eq.~\ref{eq: formularten}) with linear velocities (Eq.~\ref{eq: formularthree}) prior to inputting them into a neural network.}

{In Fig.~\ref{fig:figure0_theory}, the acceleration from our naive model is compared to the linear velocity of each galaxy in six different halo mass intervals above $2.58 \times 10^{12} h^{-1} \mathrm{M}_{\odot}$ in the simulation. In all mass intervals, the model results match the linear velocities with the same slope (red line) and small scatters, which indicates that using only the linear combination of counts and barycenters of the spherical shells can predict the linear velocities of galaxies well. }


{From the test above, we learn that a quantity involving subhalo counts and barycenters in spherical shells around a galaxy provides an accurate estimate of its peculiar velocity.} Hereafter, we will prepare this information for each target galaxy and feed it into the training network. In the data preparation, we use the thin spherical shell hypothesis, that is to say, the thickness of each spherical shell is very thin relative to the radius of the spherical shell. The detailed description about the thickness of each spherical shell is shown in Table~\ref{tab:table1}. In short, within 22 $\smpc$, 10 shells separated equally in log scale are used, while for radius greater than 22 $\smpc$, 43 shells are used to include more environment information at large scales. In total 53 spherical shells are used for the network to evaluate the gravitational effect from the nearby galaxies on a target galaxy's peculiar velocity.

The model above is assumed to be working in real space by default. But observational data is taken in redshift space. We will demonstrate later that the same model works equally well for mock data in both redshift space and real space.  Therefore, it is directly applicable in real observations. 

\subsection{Artificial neural network}

Artificial neural networks (for some reviews see \cite{Lecun2015, goodfellow2016deep}) 
are suitable for solving problems with no known specific mathematical expressions. By
constructing a nonlinear parametric model, the network converts complex problems into
non-convex optimization and optimizes the trainable parameters by gradient-descent based
methods (e.g. stochastic gradient descent \citep[][]{bottou1998online}). The process of
optimizing trainable parameters by feeding a series of data points into a fixed network
architecture is called \textit{training}. 

A standard feed-forward neural network consists of multiple layers. Each layer performs a
weighted linear combination of its inputs, followed by an element-wise nonlinear activation
function and a bias term. These weights and biases on all layers constitute the trainable
parameters of the network.

The output of each layer is the input of the next layer. For layer $n$, if we set the input vector as $\boldsymbol{x}_{n-1}$, weight matrix $\mathbfss{W}_n$ and bias vector $\boldsymbol{b}_n$, then the output of this layer is 
\begin{equation}
    \boldsymbol{x}_n = a(\mathbfss{W}_n\boldsymbol{x}_{n-1}+\boldsymbol{b}_n).
\end{equation}
The function $a$ represents a nonlinear activation function. In this paper, we use the rectified linear unit (ReLU, \citep{hahnloser2000digital,nair2010rectified}). In the training process, we need to define a loss function to represent the error between the output results and the training targets. Through the gradient-descent {based} method, the weight matrix $\mathbfss{W}_n$ and bias vector $\boldsymbol{b}_n$ are continuously optimized, and the loss function is also decreasing. 

In ANN, increasing the number of layers $N$ always expands the
capacity of the network, by enlarging the hypothesis space of solutions that
the algorithm is able to choose from, although it may lead to difficulties in
training. Thanks to the powerful capabilities of neural networks, it can be used to predict some complex nonlinear relationships. At the same time, the training process of the neural network makes it a data-driven method, which endows it with a powerful ability to utilize prior information in the data.

In this work, we use \ourModel to approximate the function $f$ in Eq.~(\ref{eq: formularnine}). For each target galaxy, we use a series of barycenters and subhalo counts of spherical shells calculated in redshift space as the input of network model. We selected a total of 53 spherical shells. The smaller the radius of the shell is, the thinner the shell we divided. The detailed configuration of the spherical shell division is shown in Table~\ref{tab:table1}. Since each spherical shell of each target galaxy has a 1D mass and a 3D barycenter, there are a total of ${53\times {4}=212}$ elements per galaxy.

We choose a five-layer fully connected neural network as our \ourModel to estimate galaxy velocity. The detailed architecture of the network is shown in Table~\ref{tab:table2}.

\begin{table*}
	\centering
	\caption{Division scheme of 53 different spherical shells in redshift space. This table lists the radial range, division scheme (log or linear space and thickness of each shell) and number of shells. For smaller radii, we use the standard of logspace to divide spherical shells, and for larger radii, we use the standard of linspace. For the spherical shell divided by linspace standard, we give the thickness of the spherical shell.}
	\label{tab:table1}
	\begin{tabular}{ccc} 
		\hline
		Radius($\smpc$) & division standard / thickness of spherical shell($\smpc$) & Number of spherical shells\\
		\hline
		1$ \sim $ 22 & Logspace & 10\\
		16$ \sim $ 60 & Linspace, 2 & 22\\
		60$ \sim $ 150 & Linspace, 10 & 9\\
		150$ \sim $ 290 & Linspace, 20 & 7\\
		290$ \sim $ 490 &Linspace, 40 & 5 \\
		\hline
		total &  & 53\\
		\hline
	\end{tabular}
\end{table*}

\begin{table*}
	\centering
	\caption{The Network structure. Our network consists of 5 fully connected layer. The output shape describes the output size of each layer. Fully connected layer 1 to fully connected layer 5 are followed by a ReLU (Nair \& Hinton 2010) activation function in our network.}

	\label{tab:table2}
	\begin{tabular}{p{3.5cm}p{3.5cm}<{\centering}p{3.5cm}<{\centering}} 
		\hline
		Layer & Output shape & Activation function \\
		\hline
		input & (batch\_size,212) & None\\
		fully connected 1 & (batch\_size,1024) & ReLU\\
		fully connected 2 & (batch\_size,512) & ReLU\\
		fully connected 3 & (batch\_size,512) & ReLU\\
		fully connected 4 & (batch\_size,64) & ReLU \\
		fully connected 5 & (batch\_size,8) & ReLU\\
		output & (batch\_size,1) & None\\
		\hline
	\end{tabular}
\end{table*}

In the training of ANN, the learning rate and mini-batch size are $5\times10^{-5}$ and $250$, respectively. We choose xavier\_Initialization \citep{glorot2010understanding} to initialize the trainable parameters, the Adam optimizer \citep{kingma2014adam} and L2 regularization \citep{hanson1988comparing}. {We find that our network can achieve convergence results after a few hours of training.}

To train the model, we need to provide the expected value of galaxy velocities to calculate the loss function. Instead of the true line-of-sight velocity, we choose the smoothed velocity $v_s$ as the expected value. To obtain the smoothed velocity $v_s$, we assign the galaxy velocities to a $250^{3}$ grid using the nearest particle (NP) \citep{Zheng2013} method, with the 4$\smpc$ side length of each cell. Then for each galaxy, we search for its nearest cell and use the corresponding $v_s$ of this cell as the $v_s$ of the galaxy. The advantage of choosing smoothed velocity instead of true velocity as the target value is that it reduces the disturbance caused by small-scale randomness of galaxy velocity in training. {The loss function selected in this paper is mean absolute error (MAE): 
\begin{equation}
    \rm {MAE}= \rm {mean} (| \boldsymbol y-\boldsymbol v_{s}|)
    \label{eq: mae},
\end{equation}
where $y$ denotes the predicted velocity of galaxies and $v_{s}$ denotes the corresponding smoothed velocities we mention above.
The MAE loss function is less affected by data outliers and is more suitable for data with large error dispersion.}

\subsection{Data set}

To train the network and for testing, a large set of high-resolution simulations is needed. In this work, we use the Indra simulations \citep{Falck2021}, a suite of $N$-body simulations evolved from 384 different random initial conditions using the L-Gadget code \citep{Springel2005}. Each simulation employs $1024^3$ dark matter particles in a periodic cubic box of comoving size 1000 $\smpc$. The cosmological parameters in these simulations are the best-fit parameters of WMAP7 \citep{Komatsu2011},
\begin{equation}
    \left\{\Omega_{\mathrm{m}}, \Omega_{\Lambda}, \Omega_{\mathrm{b}}, h, \sigma_{8}, n_{\mathrm{s}}\right\} = \left\{ 0.272,0.728,0.045, 0.704,0.81,0.967\right\},
\end{equation}
where, respectively, $\Omega_{\mathrm{m}}$, $\Omega_{\Lambda}$ and $\Omega_{\mathrm{b}}$ are the present-day density parameters for matter, cosmological constant and baryons; $h$ = $H_0$/100 (km/s/Mpc) with $H_0$ the Hubble constant, $\sigma_{8}$ is the rms linear matter density fluctuation smoothed on scales of 8 $\smpc$ at $z=0$, and $n_s$ is the primordial power spectrum index of the density perturbations.

In this work, we use 12 independent simulations at $z = 0$, where the ratio of training, validation and test sets are $3:1:8$. The validation and test sets do not participate in the process of gradient descent. The purpose of using the validation set is to quickly adjust hyper-parameters and monitor whether the model is overfitting. The test sets do not participate in the training process, nor in the fine-tuning, and they are only used for model evaluation. The
large size of the test sets is necessary to overcome cosmic variance on large scales.

For each simulation, groups and subhalos are identified through the friends-of-friends (FOF) \citep{Davis1985}) and Subfind \citep{Springel2001} algorithms. 
Taking into account the mass resolution of Indra simulation, we select subhalos with more than $40$ dark matter particles ($2.58 \times 10^{12} h^{-1} \mathrm{M}_{\odot}$). In each simulation, there are about $1.5$ million selected subhalos. Note that it is easy to extend our method to other subhalo mass thresholds by changing the selection criteria or simulation. {And when we run the \ourModel on the test set, it takes just a few minutes for the neural network 
 to output the velocities of all selected subhalos in each simulation.}

The network model is only trained by simulation data using WMAP7 cosmology. To test the dependence of our model on the underlying cosmology, we use another pair of simulations otherwise with Indra's settings, with WMAP5 \citep{Hinshaw2009} and WMAP9 \citep{Bennett2013} cosmological parameters. 


\subsection{Galaxy Formation Model--Subhalo Abundance Matching}
\label{sec:SHAM}
Subhalo abundance matching (SHAM) is an efficient way to populate galaxies in dark-matter halos, and requires minimal assumptions about galaxy formation. In this model, a galaxy is placed in dark-matter subhalos with its stellar mass (or luminosity) monotonically related to the maximum mass (or maximum circular velocity) ever attained by the subhalo during its history.  It has become a popular method for modelling the large scale distribution of galaxies, and is the most convenient way for us to populate galaxies in the Indra simulations.

In this work, we only care about the positions and velocities of massive galaxies distributed on large scales, so a simplified SHAM model is used: we populate the galaxies in every subhalo with mass greater than $2.58\times 10^{12}h^{-1}M_{\odot}$ (about 40 simulation particles). 
In our subhalo sample, more than $92\%$ are main halos. The masses of these subhalos at redshift zero are actually their maximum masses over their whole history, so the simplified version should give similar results to the usual model.

\section{RESULTS}  \label{sec:results}

In this section, we will present the results of the testing sample from our \ourModel
on the estimation of the line-of-sight velocity $v_{\mathrm{los}}$ for each galaxy, 
and the results on the distribution of galaxies before and after correcting the RSD effect. With Eq.~(\ref{eq: formulartwo}), {given $v_{\mathrm{los}}$,} the position of each galaxy in real space can be predicted. Therefore, the clustering statistics of galaxies in redshift space before and after the correction of RSD effect can be directly 
compared with that in real space. 

\subsection{Line-of-sight velocity of individual galaxies}

\begin{figure*}
	\includegraphics[width=\textwidth]{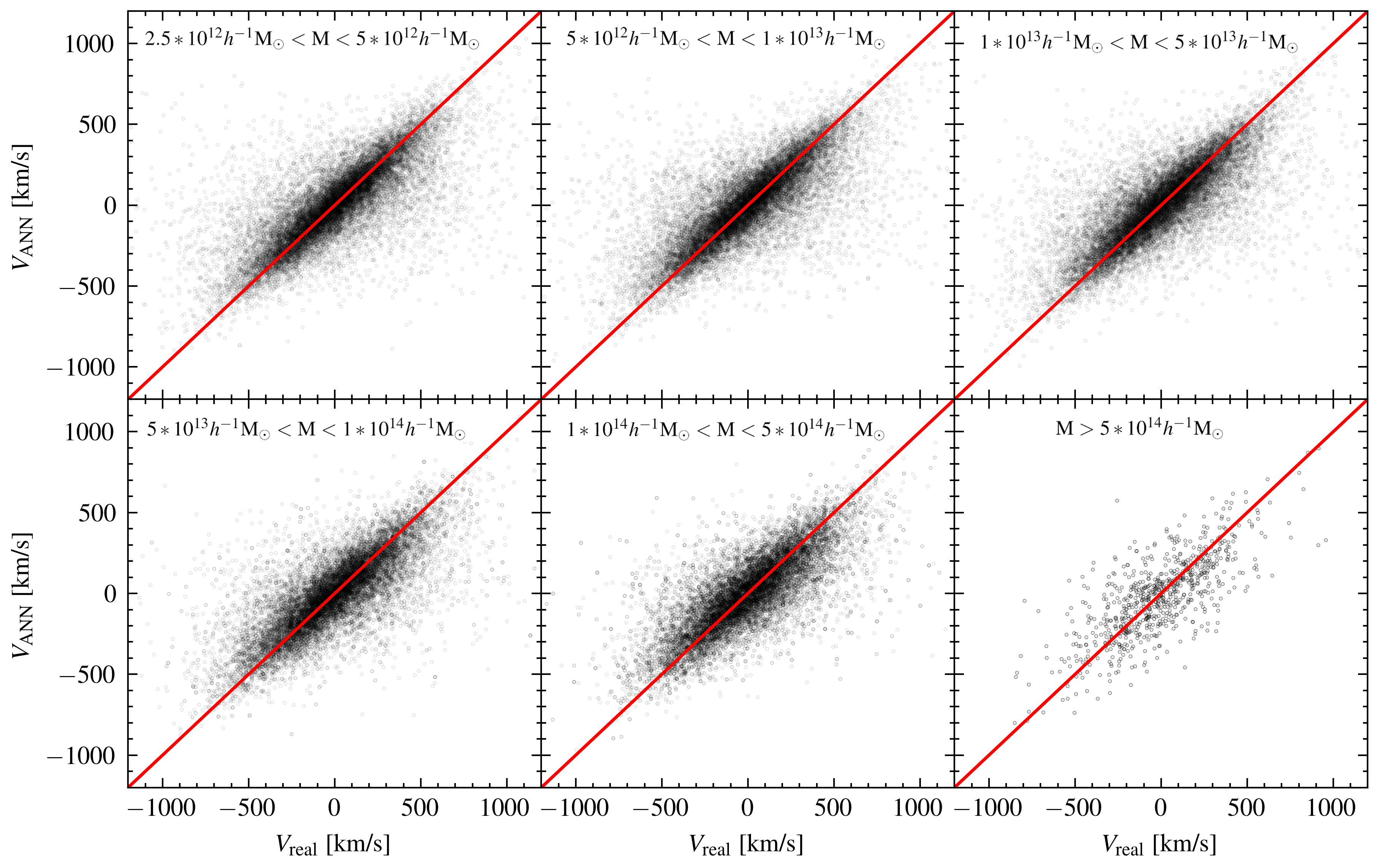}
    \caption{The predicted line-of-sight velocity versus the real velocity of each galaxy. All data points are very close to the red solid line, which indicates that most predicted velocities by \ourModel in each mass region are tightly correlated with the true velocities.}
    \label{fig:figure0}
\end{figure*}

We predict subhalo velocities with our network model. In Fig.~\ref{fig:figure0}, the 
predicted line-of-sight velocities are compared to their ground truth value in six
different halo mass intervals above $2.58 \times 10^{12} h^{-1} \mathrm{M}_{\odot}$. The red lines indicate that the predicted value is equal to the true value. In all mass intervals, the predictions match the true values very well over all velocity value ranges. Although the scatter is a bit bigger in the {most} massive mass bin, that can be explained by the sparse training and testing samples of those massive samples.

\begin{figure}
	\includegraphics[width=\columnwidth]{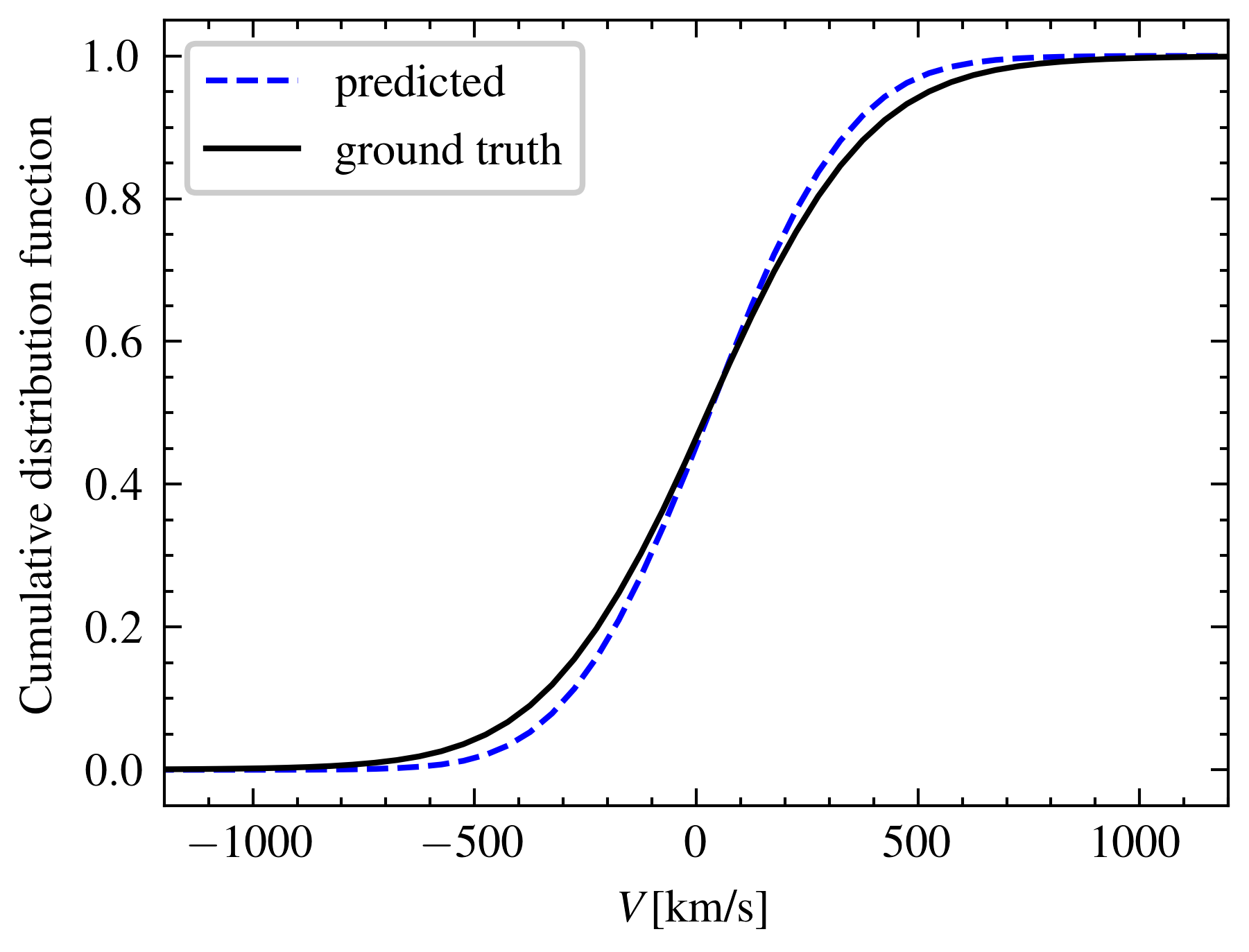}
    \caption{The cumulative distribution functions (CDFs) of the line-of-sight
    velocities of all galaxies. The black-solid and blue-dashed lines indicate
    the CDF of ground truth and \ourModel prediction individually. Since the
    proportion of galaxies with line-of-sight velocities close to $0$ is large, the
    \ourModel tends to underestimate the random part of the line-of-sight velocity that is caused by small-scale virial velocity. The steeper CDF slope of prediction at $v=0~km/s$ reflects this point. However, this will not affect RSD corrections on large scale and meso-scale.}
    \label{fig:figure0_2_new}
\end{figure}

The velocity dispersion outside the nuclear region comes from the virial velocity in halos. 
Our prediction of each subhalo velocity does not include its virial velocity,
which causes the \fog effect in redshift space. Typically, predicted subhalo velocities in the dispersion region are slightly lower than their actual velocities. 
When we check the probability distribution of the predicted and true velocities, as in Fig.~\ref{fig:figure0_2_new}, we find that the output velocity of the ANN has a large fraction close to zero. This is mainly because the variance brought by small-scale random velocities inside dark matter halos can widen the distribution of the peculiar velocity, leading to long tails. When training, the \ourModel will preferentially compress the random small-scale velocity to smaller values, so the proportion of the predicted velocity close to zero is higher in Fig.~\ref{fig:figure0_2_new} (the blue dash line). However, this will not affect the RSD correction except on very small scales (such as $s \ll 10 \smpc$). We have also established an AI model to eliminate the cluster scale \fog effect, which will be published in the next article. It is worth noting that the results of our \ourModel can obtain unbiased prediction of velocities of subhalos of different masses, without adding any velocity bias. The velocity calculated by \ourModel does not need to interpolate the velocity from the grid to the galaxy, but can directly be used to correct galaxy's RSD effect, which eliminates the error caused by interpolation.

\subsection{line-of-sight position of individual galaxy}
\begin{figure*}
	\includegraphics[width=\textwidth]{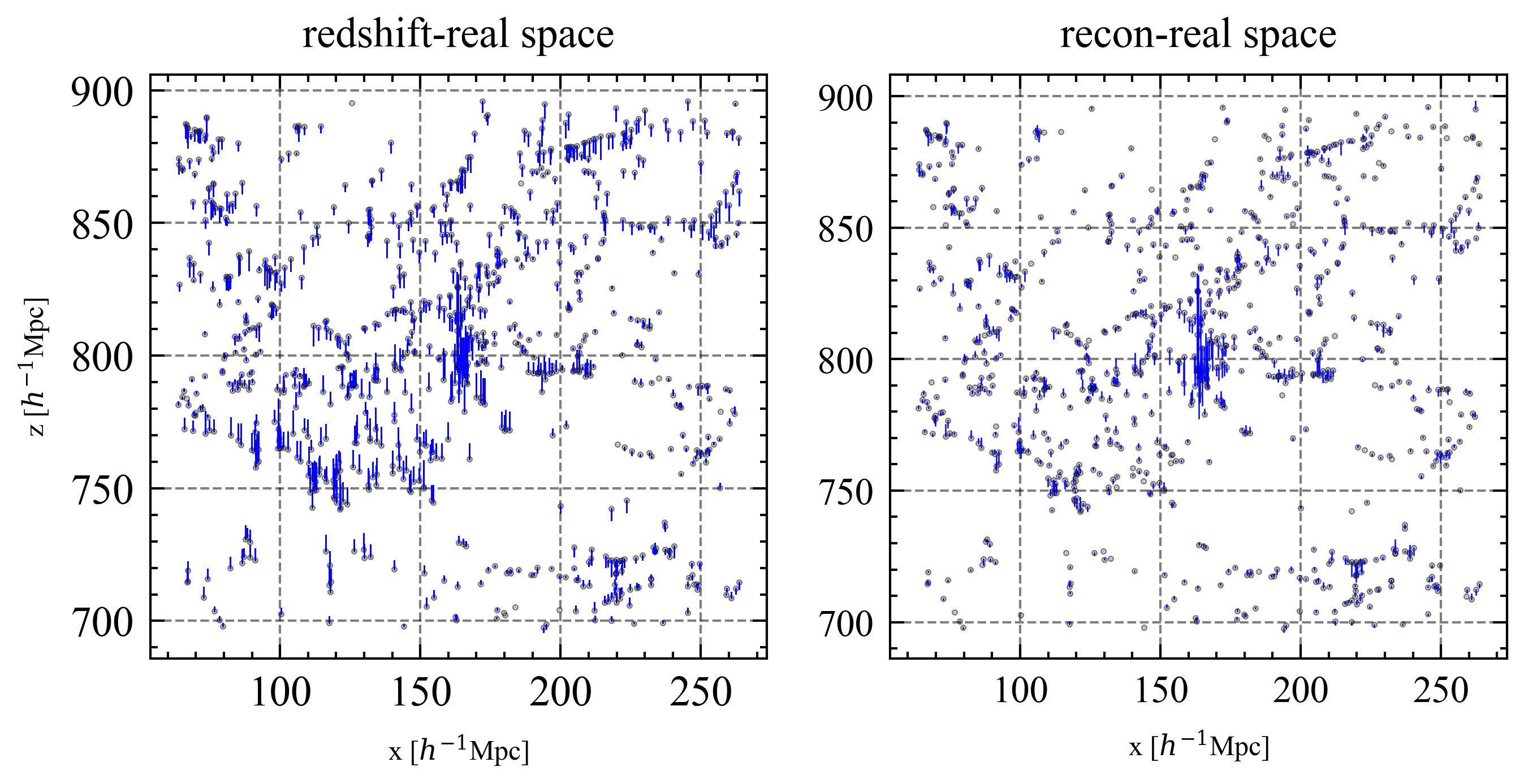}
    \caption{The galaxy displacements of redshift-real space and \rec-real space. The points in both panels denote the galaxy distribution in real space. But the blue thin lines in left panel indicate the galaxy displacements between redshift and real space, and the blue thin lines in the right indicate displacements between \rec and real space.  }
    \label{fig:displacement}
\end{figure*}

Based on Eq.~(\ref{eq: formulartwo}), the \ourModel can be used to predict the line-of-sight velocity of each individual galaxy, allowing for the estimation of their true coordinates along the line-of-sight and the reconstruction of the density field in real space. For clarity, in the following description of this paper, we refer to the galaxy distribution in real space, redshift space and after eliminating RSD effect by the network prediction as the \textbf{real}, the \textbf{distorted} and the \textbf{\rec} galaxy distribution, respectively. In Fig. \ref{fig:displacement}, we show the galaxy displacements between the real space and the \rec space by a scatter diagram. Each point represents one galaxy in a slice with depth of 12$\smpc$ in real space. The $y$-axis is the line-of-sight direction for the $z$ coordinate. The blue thin lines in left panel denote the displacements between redshift and real space, which indicates the amplitude of the RSD effect along the line-of-sight direction (y-axis). Similarly, the right panel shows the displacements between \rec and real space are drawn as blue thin lines, which indicates the residual RSD effect in \rec space. Comparing the left and right panels, the RSD effect is suppressed and most of galaxies' displacements between \rec and real space are very small. However, for dense regions where massive galaxy clusters are located, the remnants of the RSD effect are still evident. This is likely due to the loss of information when `shell-crossing' occurs in redshift space along the line-of-sight direction i.e., the relative positions of galaxies are swapped in redshift space due to strong virial motions around massive clusters. Therefore, the \ourModel does not perform very well in the regime of strong non-linearity (FoG). This may also be related to the slight disagreement between the predicted velocity CDF with the ground truth in Fig. \ref{fig:figure0_2_new}. In a separate paper, we will set up a model on peculiar velocity of galaxies in galaxy clusters to eliminate the FoG effect.

\subsection{Correlation coefficient}


After directly comparing the galaxy distribution of the real, distorted and \rec space, it will be interesting to check the clustering of galaxies after the RSD effect was eliminated. We also construct the `artificial' redshift-space galaxy distribution made by substituting our network-predicted velocity for the real velocities to produce the redshift-space field. First, we checked the correlation coefficient between the \rec and real galaxy density fields. The correlation coefficient between two fields describes their phase correlation in Fourier space. It is defined as

\begin{equation}
r(k)=P_{12}(k) / \sqrt{P_{1}(k) P_{2}(k)},
\end{equation}
where  $P_{1}(k)$ and $ P_{2}(k)$ are the auto-power spectra of the two fields, and
$P_{12}(k)$ is their cross-power spectrum. 

\begin{figure}
	\includegraphics[width=\columnwidth]{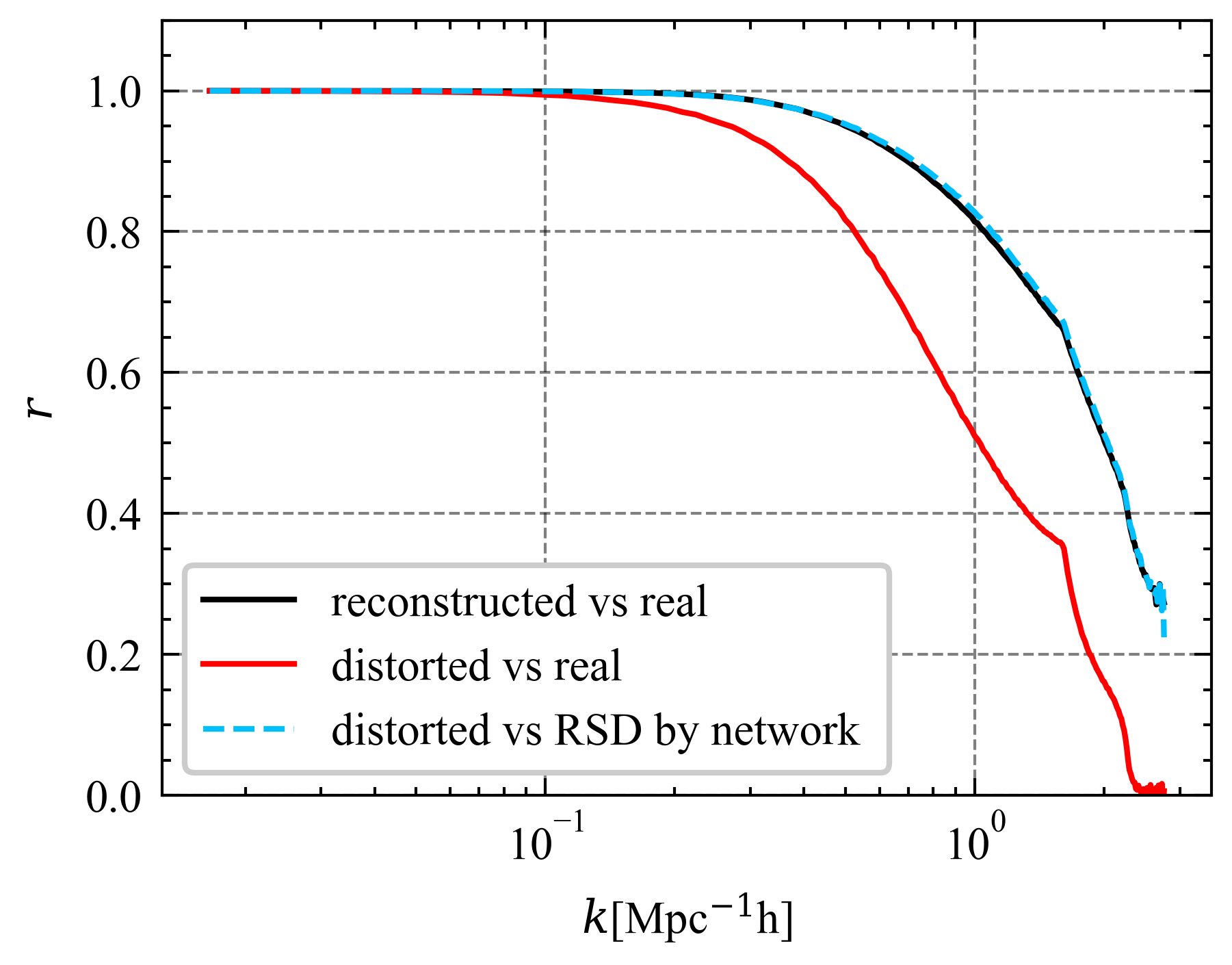}
    \caption{ The correlation coefficient.  The black solid line shows the
    correlation between the \rec and real galaxy density fields, the red line
    indicates the correlation between the distorted and real density fields, and the blue dashed line indicates the correlation between the distorted and artificial-RSD density fields made by substituting real velocity for our network-predicted velocity. The \rec density by the \ourModel is about $90\%$ correlated with the real density at $0.6 \kmpc$ and about $80\%$ at $1.0 \kmpc$. 
    }
    \label{fig:figure_corr}
\end{figure}

In Fig.~\ref{fig:figure_corr}, the correlation coefficient between the \rec and real galaxy
density fields is shown as the black solid line. As a comparison, the correlation
between the distorted and real density fields is presented as the red solid curve. The black solid curve is significantly above the red curve when $k$ is greater than $0.1 \kmpc$,
indicating the information distorted by the RSD effect was \rec effectively. Even when
red dashed curve is very close to zero around $k=2.5 \kmpc$, the \rec correlation
coefficient (black solid curve) is still above $30\%$. This remaining difference can be attributed to the FoG effect.

We also checked the correlation coefficient between the distorted and the artificial-RSD galaxy density field produced using the network-predicted velocities, which is drawn as the blue dashed line. This curve is very close to 1 when $k$ is smaller than $0.2 \kmpc$, and around $83\%$ at $k=1\kmpc$. This result indicates the velocity predicted by the network is still consistent with the true velocity at $k \leq 1\kmpc$. 

\subsection{Two-point correlation function}
\label{sec:two-point-corr}

The two-point correlation function is a very useful tool to describe the clustering of
galaxies in the Universe. In this section, we will check the clustering of galaxies in
the \rec galaxy distribution and compare it with the real galaxy distribution. It is
well-known that RSD distorts the isotropic distribution of galaxies. To test the anisotropic signal in the galaxy distribution, we check the higher-order moments of two-point correlation function by a Legendre expansion. The higher-order moment of the two-point correlation function is given by
\begin{equation}
    \xi_{\ell}(s) \equiv \frac{2 \ell+1}{2} \int_{-1}^{1} L_{\ell}(\mu) \xi(\mu, s) \mathrm{d} \mu.
\end{equation}
Here, the $\xi(\mu, s)$ is the two-dimensional correlation function, and $\mu$ is the
cosine of the angle between the redshift-space coordinate $s$ and the line-of-sight
direction. $L_{\ell}(\mu)$ is the Legendre polynomial with order $\ell$. 
In this work we look at the first two multipoles, the monopole ($\ell=0$) and quadrupole ($\ell=2$), performing the measurement with the Landy-Szalay estimator \citep{Landy1993}.

\begin{figure}
	\includegraphics[width=\columnwidth]{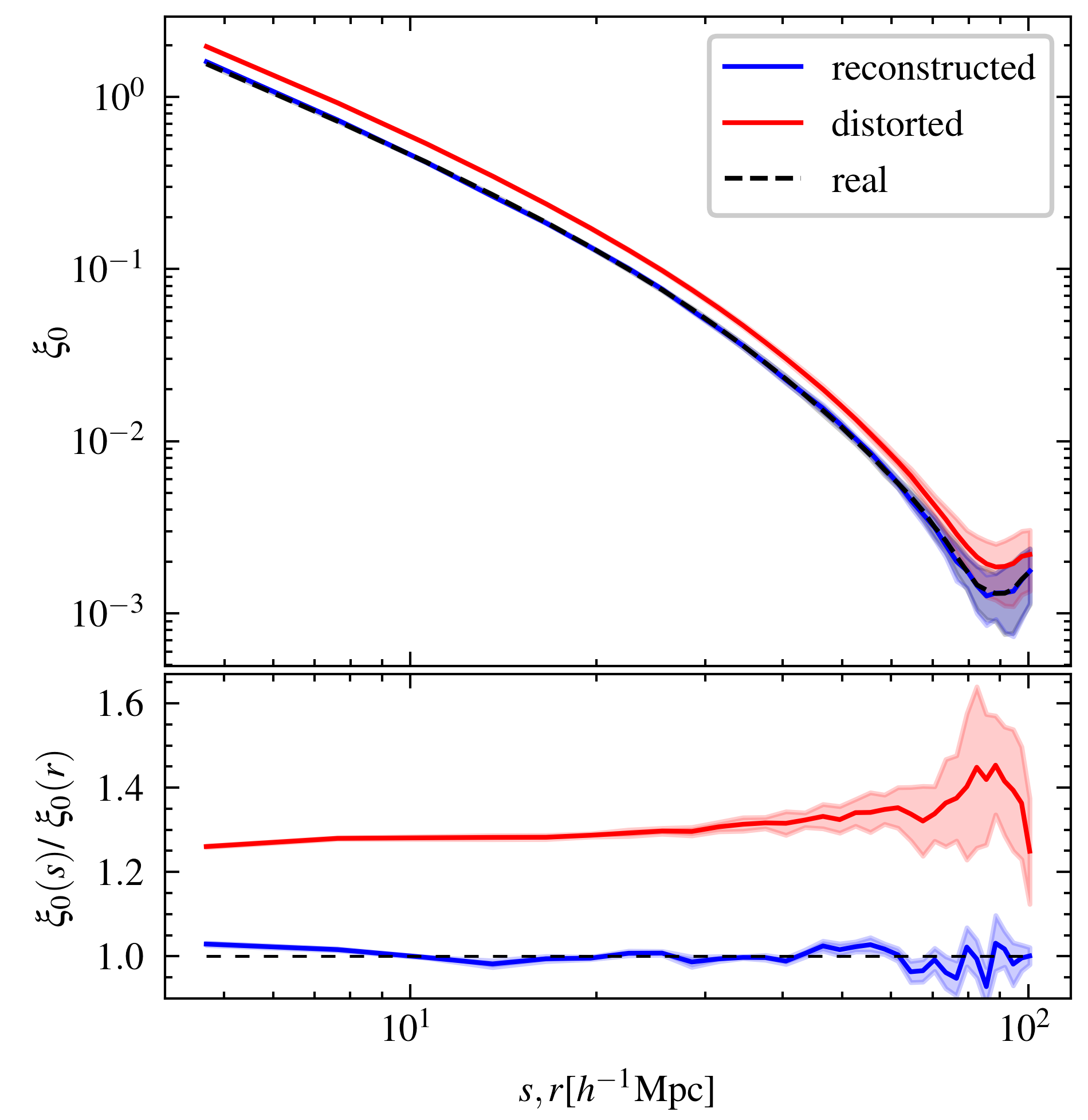}
    \caption{The monopoles of the two-point correlation functions. The blue solid, red solid and black dash lines indicate the two-point correlation functions of the \rec, distorted and real galaxy distribution at $z = 0$ respectively. The shadow areas
    represent the range of one sigma error. In the lower panel, all lines are divided by the two-point correlation function of the real galaxy distribution. Our \rec result is very close to the fiducial value 1 on scales $s > 8 \smpc$, which means the redshift distortion has been well recovered on these scales. }
    \label{fig:figure1}
\end{figure}

\begin{figure}
	\includegraphics[width=\columnwidth]{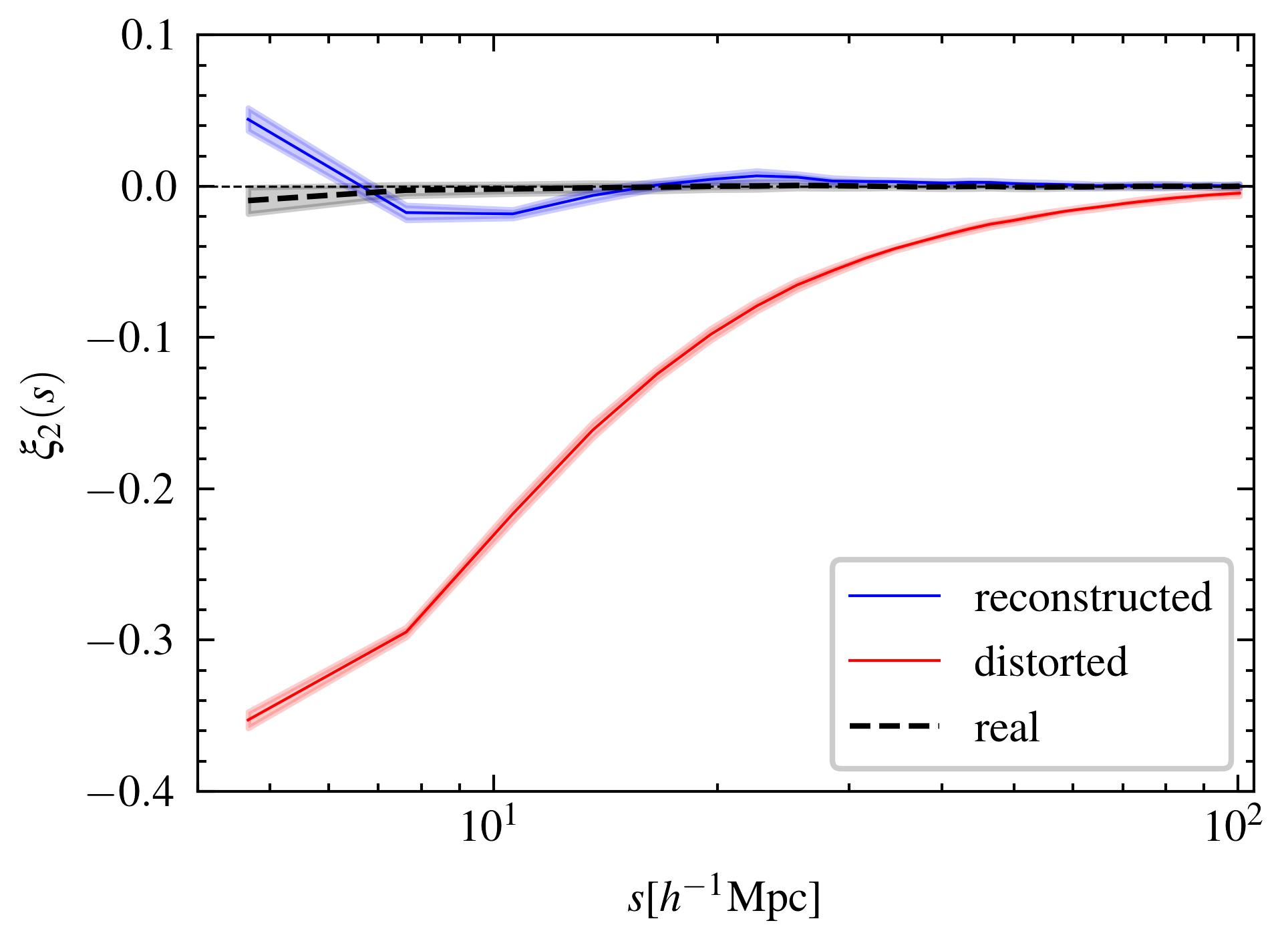}
    \caption{The quadrupole moment of the two-point correlation function.   The blue, red and black curves represent the quadrupole moment of 
    the \rec, distorted and real galaxy distributions at $z = 0$, respectively, and the shadow areas represent the corresponding one sigma error ranges. The quadrupole moment close to zero indicates spatial isotropy. Our result shows that the anisotropy caused by the RSD effect can be removed through \ourModel on scales $s > 10 \smpc$ reliably.}
    \label{fig:figure2_newversion}
\end{figure}

In the upper panel of Fig.~\ref{fig:figure1}, the two-point correlation function monopoles of
the real and distorted galaxy distributions are shown by black and red curves, while
the result from the \rec galaxy distribution is shown as blue curve. Here the the average value of $8$ testing samples are presented, and the shaded area indicates $1 \sigma$
scatter of these samples. The curves turn
up around 100 $\smpc$ because of the BAO peak. 
The \rec distribution result agrees with the real distribution very well over all scales. In the bottom panel, the ratio between the red/blue solid curves and the black dash
curve are presented. The \rec distribution result is very close to the real distribution with an error less than one percent on scales larger than 8 $\smpc$. On smaller scales, the difference is slightly larger, but remains less than $5$ percent. 

In order to check if the \rec galaxy distribution is isotropic or not, in Fig.~\ref{fig:figure2_newversion}, the quadrupole moment of the two-point correlation
function of galaxies, $\xi_2(s)$, is presented.  Same as Fig.~\ref{fig:figure1}, the results of the real, distorted, and the
\rec distributions are indicated by black, red, and blue curves respectively, and the shaded ranges indicate the $1 \sigma$ error for the eight testing samples. On large
scales, because of the RSD effect, the quadrupole moment in redshift space departs
from zero at around $100 \smpc$, and the difference gets larger towards smaller scales. For scales below $10 \smpc$, the deviation also shows up in the result for the real space due to the non-linear evolution. In the \rec galaxy distribution, the anisotropic signal at large scales almost completely disappears. As
expected, the result of the \rec distribution is very close to that in real space.


\subsection{Power spectrum}
\label{sec:power-spectrum}

\begin{figure}
	\includegraphics[width=\columnwidth]{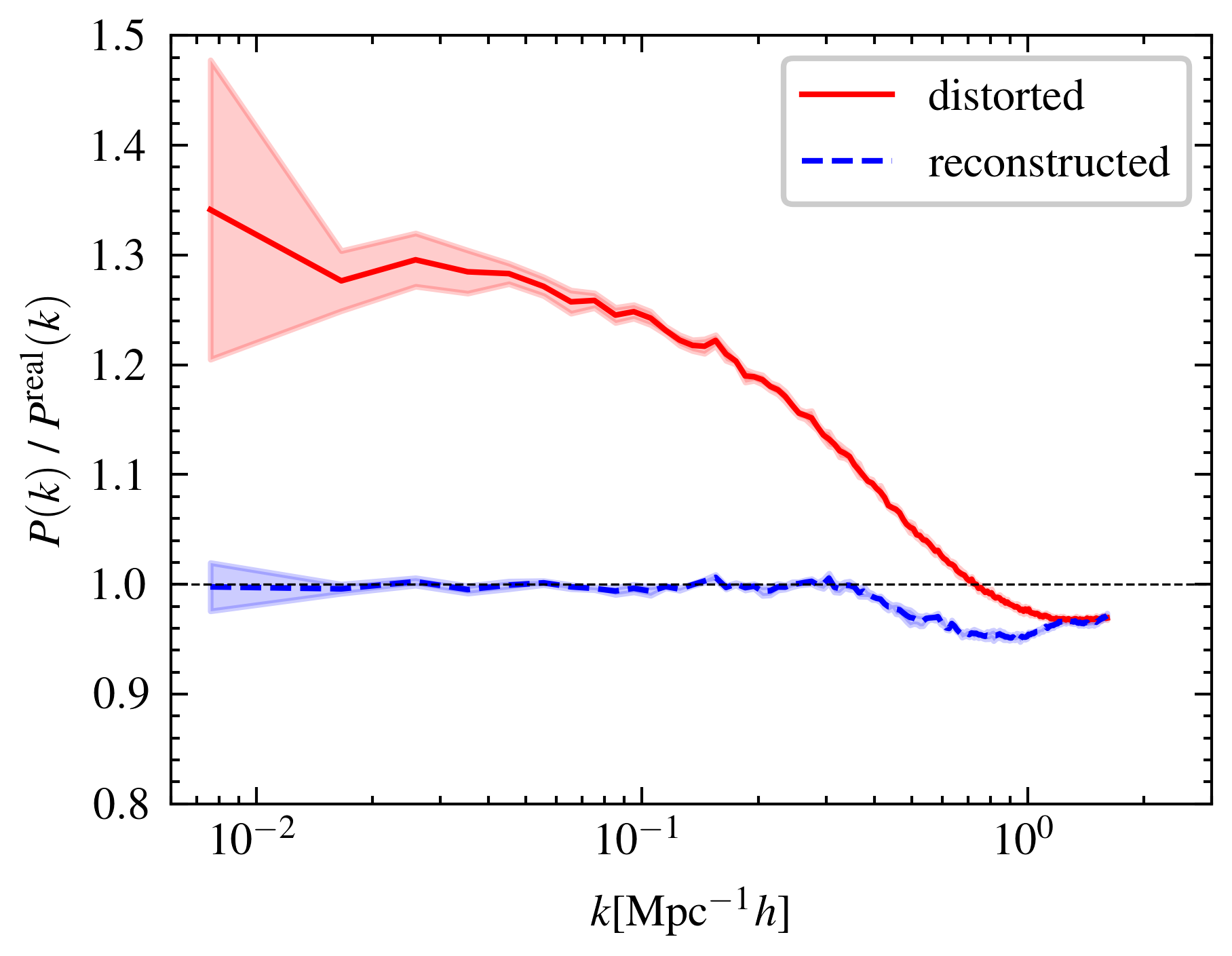}
    \caption{The power spectrum monopoles. This picture shows the ratios of the
    power spectrum monopoles of the distorted and \rec distortion to that of the
    real galaxy distribution at $z = 0$, by red solid and blue dashed curves separately. The shadow area represents the range of one sigma error from 8 independent realizations of the Indra simulation. We find the error of the \rec result is less than $3\%$ on the scales of $k < 0.5 \kmpc $, and less than $5\%$ over the full range of scales. The large-scale power spectrum dispersion caused by the cosmic variance is also compressed well.}
    \label{fig:figure3}
\end{figure}

\begin{figure}
	\includegraphics[width=\columnwidth]{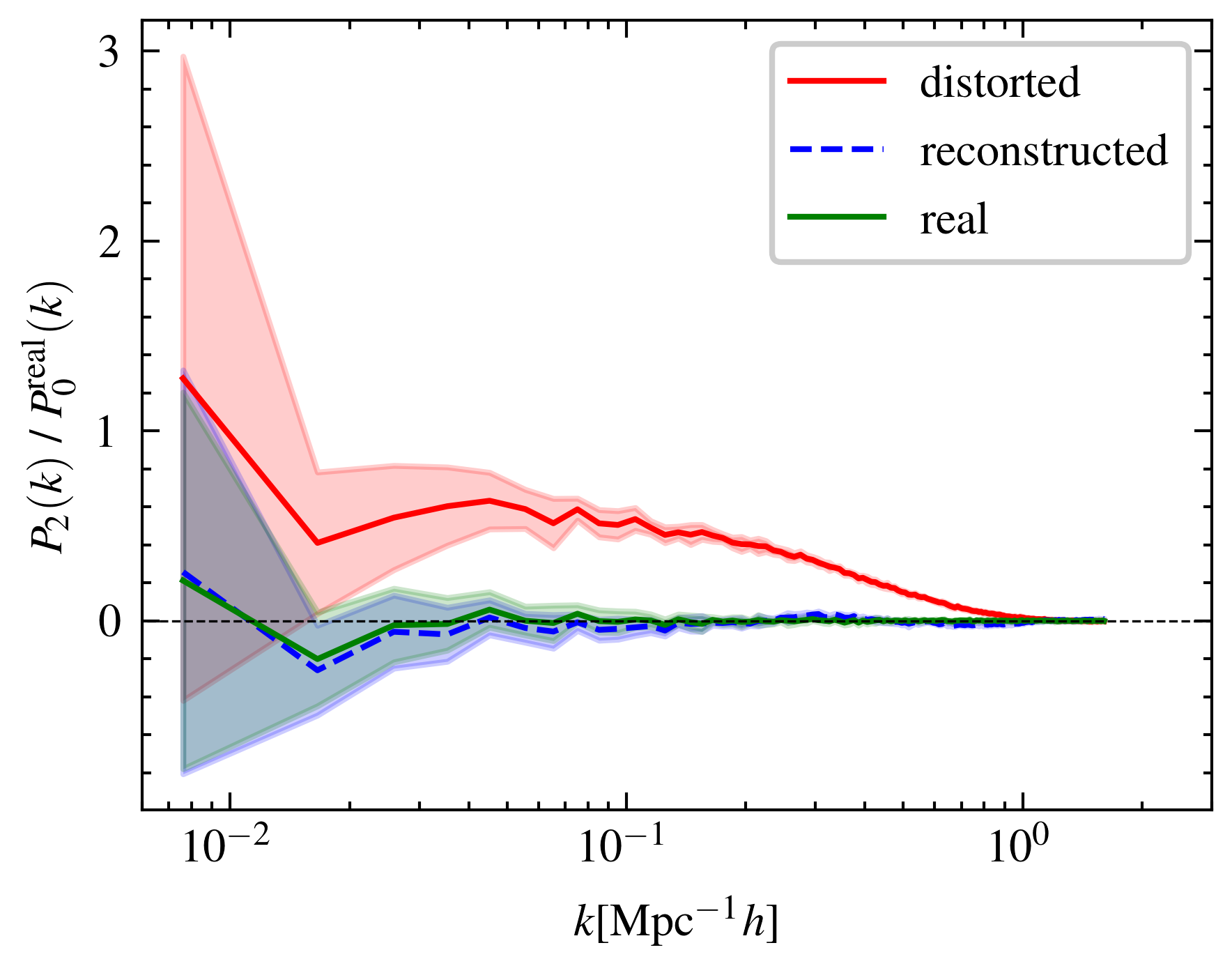}
    \caption{The quadrupole moment of the galaxy power spectrum, which is normalized by dividing the momopole moment.  The red solid, green solid and blue dashed lines indicate the quadrupole moments of the power spectrum in the distorted, real and \rec distribution at z = 0 respectively, and the shaded area represents the corresponding one sigma error range. The quadrupole moment in the corrected distribution is almost equal to the real distribution at intermediate and small scales. On large scales, the one $\sigma$ scatter of the blue line is smaller than that of the real distribution, and the slight offset of the two is likely due to the large-scale cosmic variance.}
    \label{fig:figure3_2}
\end{figure}

In Fig.~\ref{fig:figure3}, the ratios of the monopoles of the power spectra in the distorted galaxy distribution and the \rec galaxy distribution to the real galaxy distribution are presented by red solid and blue dashed curves separately, and the shaded areas indicate $1 \sigma$ scatter of eight samples. 
The curves upturn at scales $k > 1\kmpc$ due to assignment error, while the RSD effect uplifts the power spectrum almost over all scales when $k < 1 \kmpc$. After the correction, the RSD effect is suppressed almost completely, with an error less than $3\%$ at scales $k < 0.5 \kmpc$, corresponding to $12 \smpc$, with an error less than $5\%$ over the whole range of scales. Furthermore, the scatter for the \rec results is less than $2.5\%$, indicating the error bar of our RSD modelling is very small. These results echo that both large-scale and
small-scale redshift distortions have been accurately modelled. 

The Legendre expansion is also used to obtain the quadrupole moment of the power
spectrum, and the results are shown in Fig.~\ref{fig:figure3_2}. Here the quadrupole
moments are normalized by their corresponding momopole moments. The results for the distorted, real and the \rec distribution are shown as red, green and blue
curves separately, and the shaded areas indicate the $1 \sigma$ error for the eight
testing samples. As shown in Fi.g~\ref{fig:figure3_2}, the quadrupole
moment is always above zero, at almost all scales in redshift space, and gets bigger at larger scales. With our correction based on \ourModel, the quadropole moment is suppressed and gets very close to the real distribution over the entire $k$ range. At large scales, $k < 0.1 \kmpc$, the scatter is large on these scales due to the cosmic variance, but substantially that that of the distorted galaxy distrbution. The results indicate the spatial anisotropy due to the RSD effect in the
galaxy distribution has been suppressed efficiently by our method.

\section{Applications and Discussions} \label{sec:discussion}
With the line-of-sight velocity accurately determined by the network model, not only
the RSD effect can be well modelled and eliminated in the statistics of galaxy
clustering as we presented in Sections \ref{sec:two-point-corr} and
\ref{sec:power-spectrum}, but our method is useful for estimating the full three dimensional cosmic density and velocity fields. Accurate estimation of line-of-sight velocities is widely applicable in astronomy and cosmology, for: BAO
reconstruction \citep{Mao2021}, the Alcock-Paczynski test \citep{Alcock1979}, the kinetic Sunyaev–Zeldovich (kSZ) effect \citep{Sunyaev1972,Sunyaev1980},  and so
on. In this section we will briefly introduce the application of our results on the
cosmic velocity reconstruction, study the cosmology dependence of our model and discuss the small-scale effect of our results.

\subsection{Cosmic velocity field}
Determination of the cosmic velocity field is always a big challenge, especially the
velocity along the line-of-sight direction, which is contaminated by the galaxy's recession velocity. Here we only focus on the line-of-sight direction, and the velocity
perpendicular to the line-of-sight could be determined by the cosmic density with
Eq.~(\ref{eq: formularthree}) directly.

\begin{figure*}
\includegraphics[width=\textwidth]{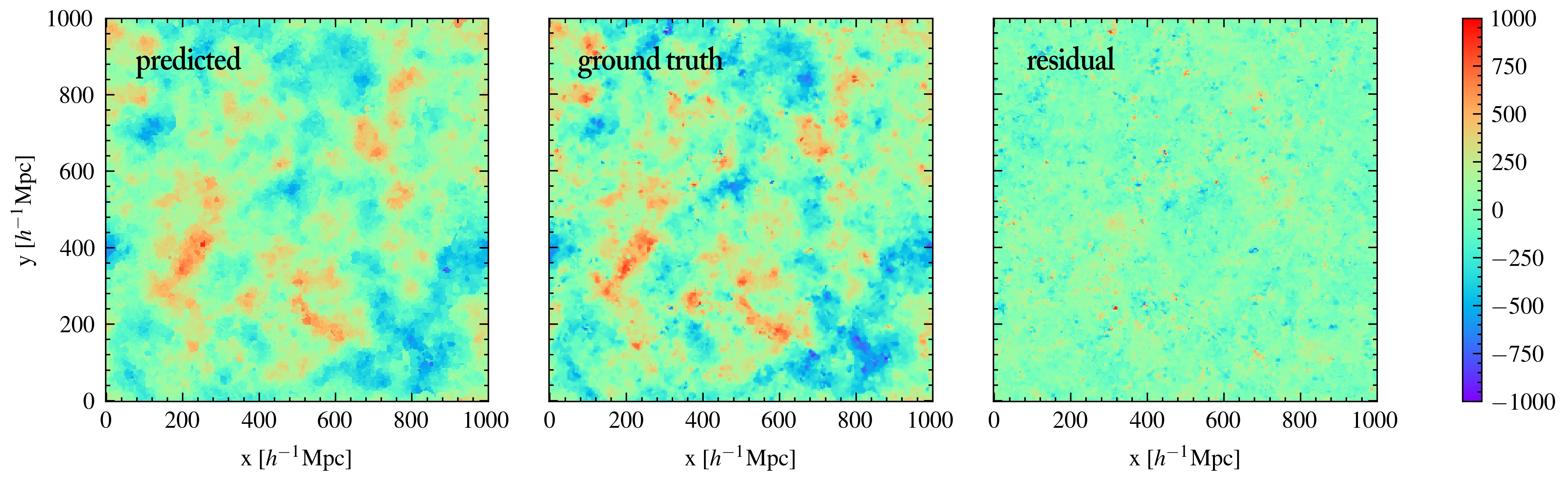}
\caption{Velocity maps along the line of sight in a slice with depth of $4 \smpc$. From left to the right, the velocity maps for the predicted and ground-truth galaxy distributions, and their residual are presented respectively. Since the RSD effect distorts galaxies position, we show the predicted velocity map based on the corrected galaxy distribution. The predicted map converges with the ground truth very well.
}
\label{fig:figure9}
\end{figure*}

To get the line-of-sight velocity at any given position in the universe, the nearest-particle (NP) scheme \citep{Zheng2013} is used to interpolate the velocity of the nearest target galaxies to that in any positions. In Fig.~\ref{fig:figure9}, velocity maps along the line of sight (in the $z$ direction, perpendicular to the slice) in a slice with depth of $4 \smpc$ in our testing sample are presented. From left to the right, the velocity maps for the predicted, ground truth, and their difference are presented respectively. Note that the slice of predicted velocity is based on \rec galaxy distribution and the ground truth velocity maps is based on real galaxy distribution. 
Generally, the predicted map agrees with the ground truth very well. Because no smoothing is adopted, there are some unexpected very large absolute values of velocity in some pixels of the ground truth map, and hence the residual map. As we only model the line-of-sight velocity at large scales in this work, the large peculiar velocity of some galaxies caused by the non-linear evolution of dark halos is not treated carefully and should be responsible for these maximum value pixels. In a subsequent paper we will focus on eliminating this effect.

\begin{figure}
\includegraphics[width=\columnwidth]{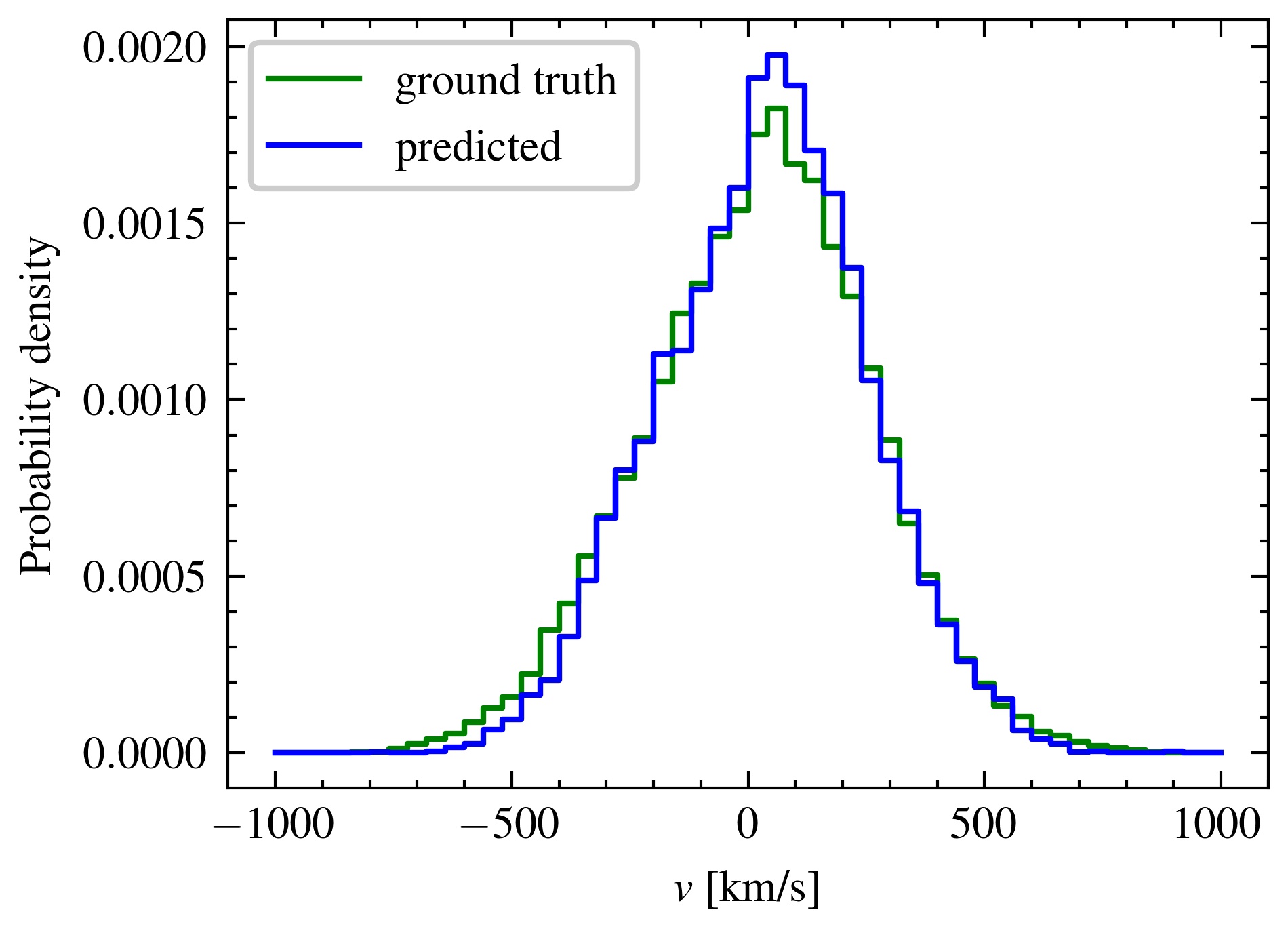}
\caption{The probability density function of velocity fields in the same $4 \smpc$ slice as used in Fig.~\ref{fig:figure9}.  The green solid line is Pthe DF of the real-space galaxy velocity field, and the blue solid line is the PDF of velocity field predicted by ANN.}
\label{fig:figure10}
\end{figure}

Furthermore, in Fig.~\ref{fig:figure10}, the probability density functions (PDF) of the cosmic velocity alone the line-of-sight for the same slices in Fig.~\ref{fig:figure9} are presented. The green and blue solid curves indicate the results for the ground truth and our predicted field respectively. Both distributions actually agree with each other, except that the PDF of the ground truth is a little higher than the predicted one at the two ends, for the same reason of those
high value of pixels in Fig.~\ref{fig:figure9}.

\subsection{Dependence on galaxy formation model}
In section \ref{sec:SHAM}, a SHAM is employed to distribute galaxies within dark-matter halos: each subhalo with a mass exceeding $2.58\times 10^{12}h^{-1}M_{\odot}$ contains one galaxy. This simple model prompts the inquiry of whether the trained network can be applied to other galaxy catalogues produced by various galaxy formation models, particularly those that incorporate more advanced physical processes in galaxy formation and evolution.

In this subsection, we use the galaxy catalogue from the Millennium-I simulation \citep{Springel2005} with the semi-analytic galaxy formation model of Guo11 \citep{Guo2011}, which is widely used for studies of galaxy formation and evolution. The size of the simulation box is 500 $\smpc$ and the mass of simulation particle is $8.6\times 10^{8}h^{-1}M_{\odot}$. A WMAP1 cosmology with $\{\Omega_{\mathrm{m}}, \Omega_{\Lambda}, \Omega_{\mathrm{b}}, h, \sigma_{8}, n_{\mathrm{s}}\} = \{ 0.25,0.75,0.045, 0.73,0.9,1\}$ is adopted, which is slightly different from the training set. Galaxies with a total mass greater than $2.6\times 10^{12}h^{-1}M_{\odot}$ at $z=0$ are finally selected as the test sample.

\begin{figure*}
	\includegraphics[width=\textwidth]{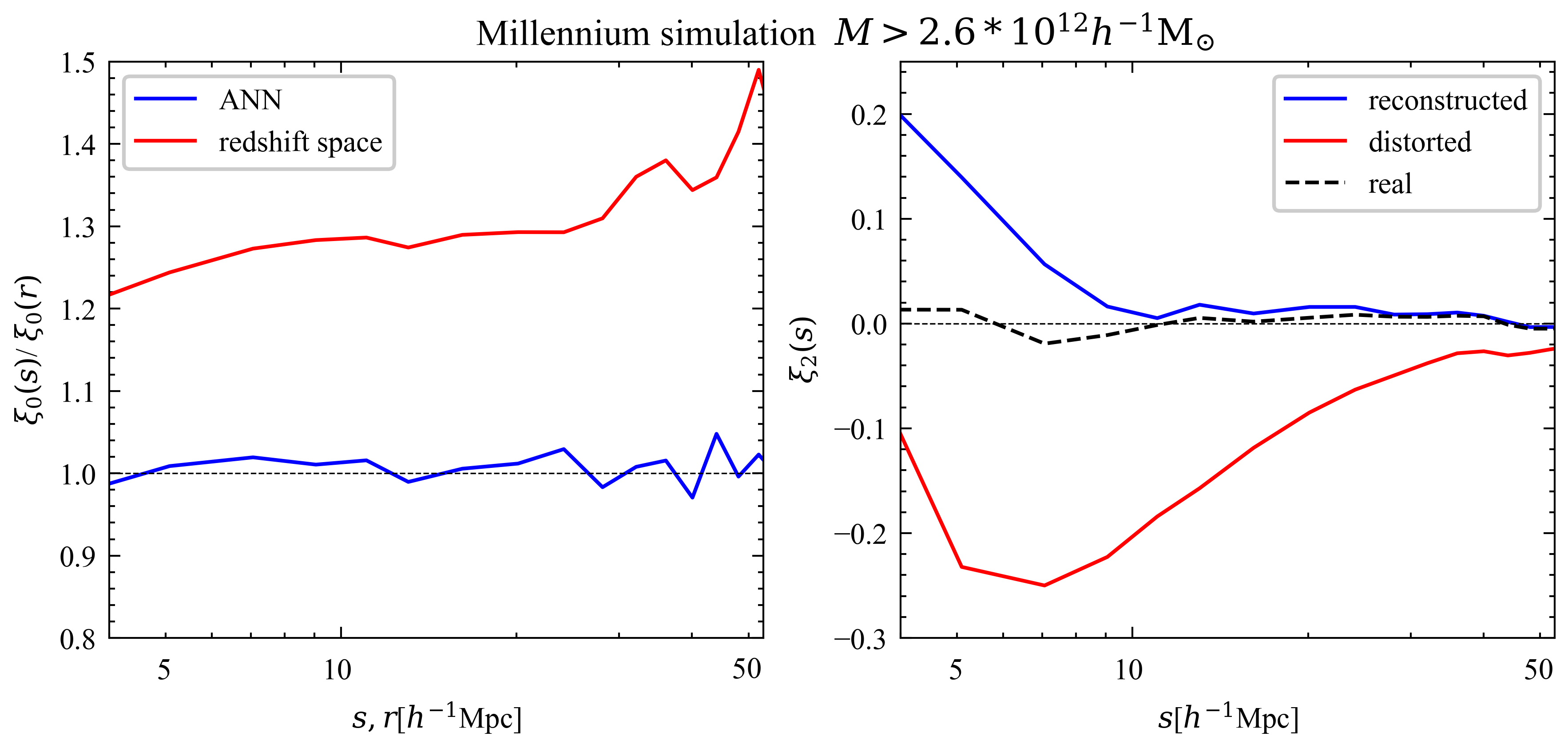}
    
    \caption{{The monopoles and quadrupoles of the two-point correlation functions in  Millennium galaxy catalogues. The left panel of the figure displays the monopole moment of the distorted and \rec distributions, that are divided by that of the real galaxy distribution in the Millennium simulation, represented by red and blue solid curves respectively. The right panel shows the quadrupole moment of the distorted, \rec and real space two-point correlation function. }}
    \label{fig:millen1}
\end{figure*}

The monopoles and quadrupoles of the two-point correlation functions in the Millennium galaxy catalogues are shown in Fig.~\ref{fig:millen1}. The left panel of the figure displays the ratios of the monopole moment of the distorted and \rec distributions to that of the real galaxy distribution, represented by red and blue solid curves respectively. The right panel shows the quadrupole moment of the distorted, \rec and real space two-point correlation function. For the monopoles, the reconstructed distribution result agrees with the real distribution very well over all scales. For the quadrupoles, the
anisotropic signal at large scales almost completely disappears at $s>8\smpc$, as well as what the networks do for the SHAM galaxy catalogues.

{In the above tests, galaxy catalogues with different galaxy formation models or different cosmological models are fed into the network, the RSD effect is still precisely eliminated, and the spatial anisotropy caused by RSD is removed. This indicates that our network can be applied to galaxy distributions that differ from the training sets, and potentially to the real observational data in the future.}

\subsection{Cosmology dependence in WMAP family models} 
In this work, the training of the \ourModel is based on the simulation data in a specific set of cosmological parameters -- WMAP 7 years result \citep{Komatsu2011}. Therefore, testing our model in other testing data with different cosmology model is necessary. For a strict valuation like \cite{Mao2021}, we choose 2 different cosmology models: WMAP 5 years \citep{Hinshaw2009}, WMAP 9 years \citep{Bennett2013}. In the next subsection, we also test the higher bias and higher redshift WMAP 7 years model (at $z=2.0$) as an extreme test, since the difference of the parameters in different WMAP family models is small. 


\begin{figure*}
\includegraphics[width=\textwidth]{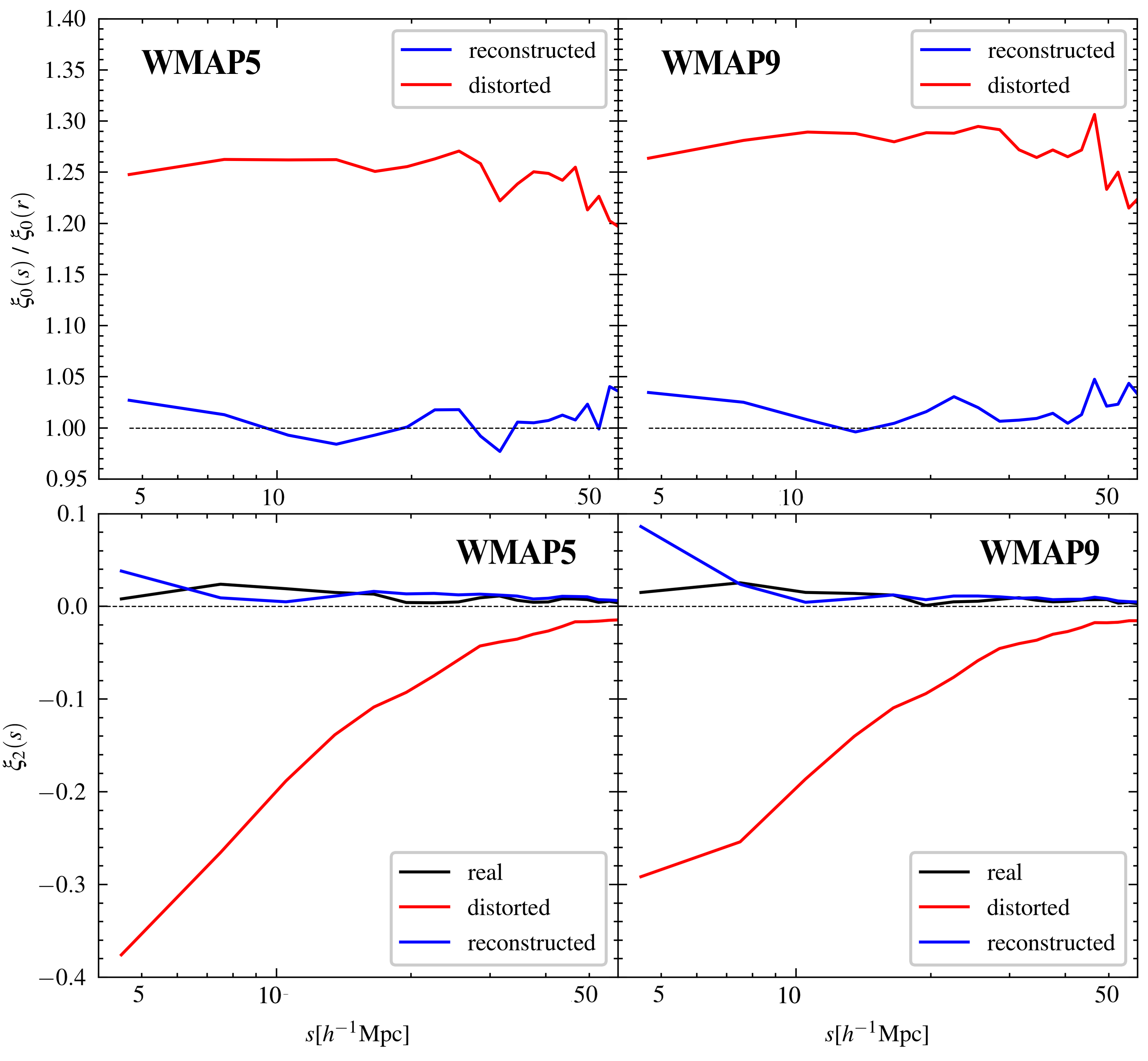}
\caption{The monopoles and quadrupoles of the two-point correlation functions in the WMAP5 and WMAP9 cosmologies. The upper left panel shows the monopoles result of WMAP5 and the upper right panel is monopoles for WMAP9. The lower two panel shows the corresponding quadrupoles results. The blue solid lines indicate the result of the \rec galaxy distribution under the WMAP5 or WMAP9 cosmologies by our \ourModel (trained by WMAP7 cosmology), while the black solid and red solid lines indicate the two-point correlation function of real and distorted galaxy distributions with different cosmological parameters. We find the cosmological parameters dependence is weak in different universe, and the anisotropy caused by RSD is basically eliminated. }
\label{fig:figure11}
\end{figure*}


In Fig.~\ref{fig:figure11}, the monopole moments of the two-point correlation functions in the WMAP5 and WMAP9 cosmologies are shown in upper left and upper right panels respectively. As in the bottom panel of Fig.~\ref{fig:figure1}, in each panel, the results from  distorted and \rec galaxy distributions are divided by that of the real distribution, and shown by red and blue curves respectively. In both panels, the blue curves are close to unity over all scales with a difference less than 5\%. In the right panel for the WMAP9 model, the blue curve seems always above unity, but the difference is still less than 5\%, and this is reasonably  good given the strong Poisson noise.

In the lower two panels of Fig.~\ref{fig:figure11}, the cosmology dependence of the quadrupole moments of the
two-point correlation function is checked. The results for distorted, real, and \rec galaxy distributions are indicated by red, black and blue curves individually. For the WMAP5 and WMAP9 cosmological models, after correction RSD using the
line-of-sight velocity, the quadrupole moment in  \rec (blue) space raises up from redshift space (red), and agrees with that in real space (black) very well.  The anisotropy caused by RSD is basically eliminated.

A caveat of the above test is that the key parameter, $f/b$ which connects the density field to the velocity field, are very similar between the fiducial cosmology and the other two WMAP models. It is perhaps not surprising that the \ourModel performs well. To stress-test the \ourModel, we create situations where $f/b$ is significantly different from the fiducial value by using simulations at high redshift, and selecting galaxies with higher biases. These are presented in the following sub-section.

\subsection{Network performance at high redshift \& massive galaxies} \label{sec:z20}

\begin{figure*}
	\includegraphics[width=\textwidth]{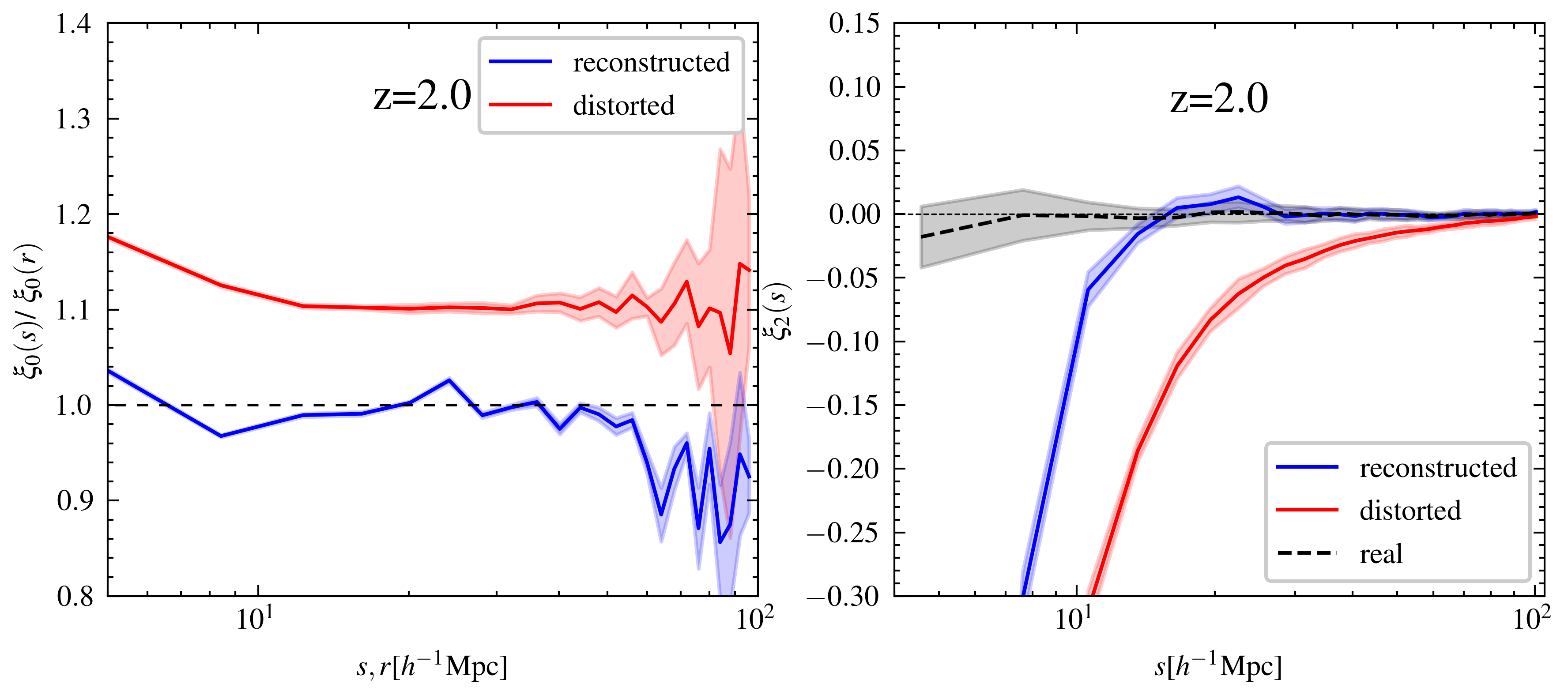}
    \caption{The left panel of the figure displays the ratios of the monopole moment of the distorted and \rec distributions to that of the real galaxy distribution at $z = 2.0$, represented by red and blue solid curves respectively. The right panel shows the quadrupole moment of the distorted, \rec and real space two-point correlation function, with the shaded area representing the corresponding one sigma error range. }
    \label{fig:z20}
\end{figure*}

At high redshift, linear theory predicts that the RSD effect will be significantly different from the training data at redshift zero due to the different bias parameter and growth rate parameter. At $z=2.06$, the growth rate is $f=0.949$ and the linear galaxy bias is $b=3.828$, which are significantly different from the values at redshift zero ($f =0.470,b=1.165$). So $\beta$ ($\beta=f/b$) is 0.248 at $z=2.06$ and 0.403 at $z=0$. The linear RSD effect is expected to weaker by approximately 60\%.
The monopoles and quadrupoles of the two-point correlation function are displayed in Fig.~\ref{fig:z20}. The left panel shows the ratio of the monopoles from the distorted and \rec galaxy distributions to those of the real distribution, represented by red and blue curves respectively. The shaded area indicates the $1 \sigma$ scatter of the ratio of 8 simulations. 

The network performance is satisfactory, despite the distortion effect is significantly different between the training sample at $z=0$ and this test sample. At $10<s<60 \smpc$, the monopole is recovered to be within 4\%. The accuracy is approximately 10\% for larger scales. It should be noted that the cosmic variance is included here since the fraction from the same field is shown. 
The right panel shows the quadrupoles of real, distorted, and \rec distributions, represented by black dashed, red solid, and blue solid curves respectively. At scales larger than $15 \smpc$, the deviation between \rec and the real distribution is very small, and the anisotropic signal at large scales is successfully eliminated. At small scales ($s<10 \smpc$), the errors for the qudrupole are much larger . This is likely because the nonlinear effect evolves differently from linear growth, and exhibit very different FoG effect at $z=2.06$, but the network is mainly successful in the mid-nonlinear regime.

\begin{figure*}
	\includegraphics[width=\textwidth]{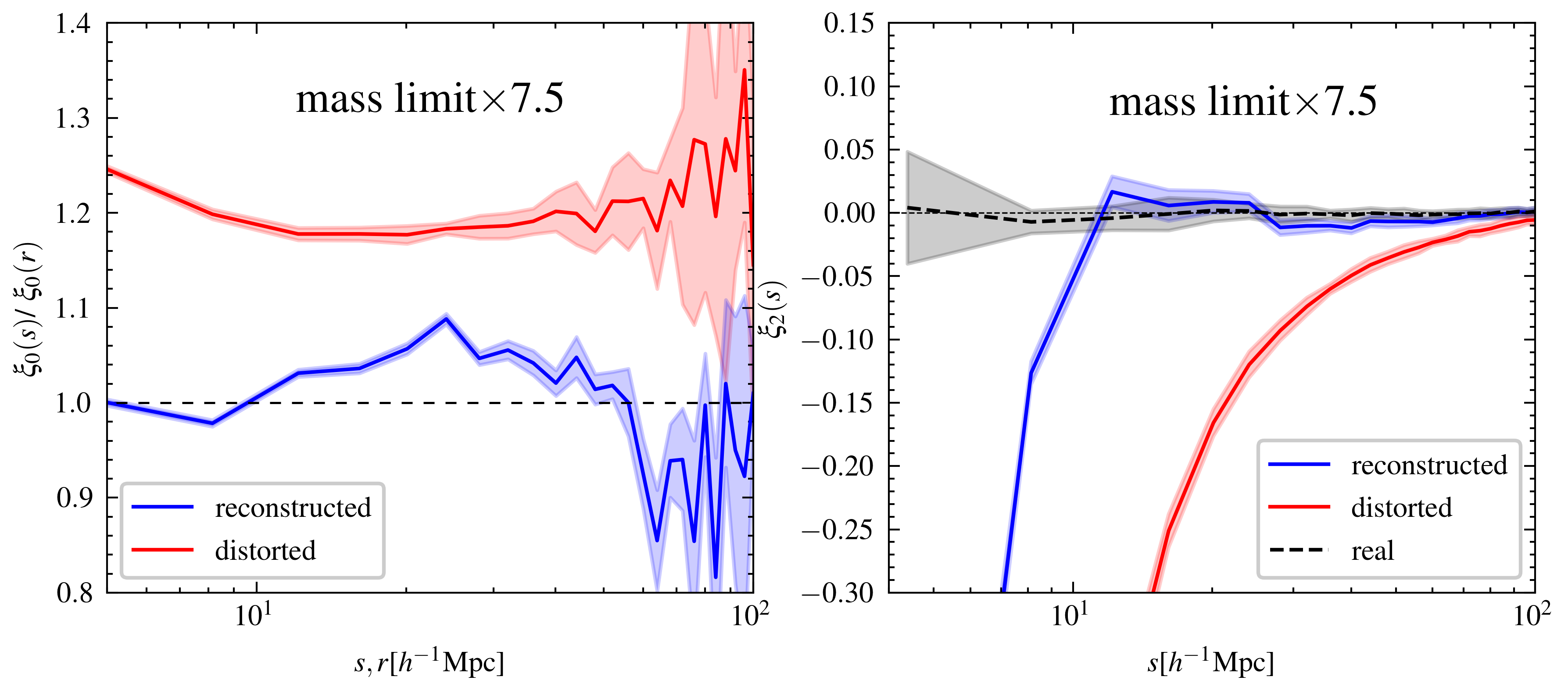}
    \caption{The monopole and quadrupole moment of the two-point correlation function for massive galaxy sample {at $z=0$}.This picture is similar to Fig.~\ref{fig:z20}.}
    \label{fig:highbias}
\end{figure*}

We see similar performance of the trained \ourModel on a massive galaxy sample with a mass greater than $1.94 \times 10^{13} h^{-1} \mathrm{M}_{\odot}$, which is about 7.5 times higher than the sample used in Section \ref{sec:results}, {at $z=0$}. The galaxy bias $b=1.806$, which is 1.55 times different from our fiducial sample. The results are shown in Fig.~\ref{fig:highbias}. We can see again that the \ourModel succeeded in removing RSD at scales greater than $\sim 10 \smpc$, but failed at smaller scales.


Therefore, despite the significant difference in $\beta$, our trained network is successful on large scales to cope with this extreme case at $z=2.06$, and for highly biased galaxy samples. This is reassuring as the network is able to build the connection between the density and velocity field independent of the $\beta$ value. This suggests that our network can be applied to various galaxy samples without needing to be retrained, and is likely to be successful when implemented with real observational data in the future.



\section{Conclusions}  \label{sec:conclusion}

In this paper, we present a new method based on artificial neural networks to predict the peculiar velocity of individual galaxies from redshift space directly, and use this velocity to obtain the real-space coordinates of the galaxies and remove redshift-space distortions.
We report its first result in simulation and obtain the unbiased and consistent result between the predicted and real velocity of each galaxy. To demonstrate the accuracy of correcting redshift-space distortions using this method, we have compared the two-point correlation functions and power spectra of the real, distorted, and \rec galaxy distributions. 
Our \rec galaxy distribution has
results consistent with the ground-truth distribution at distances $s > 8$ $\smpc$. It deviates by only 4\% at $s$ = 5 $\smpc$ compared to the real-space galaxy two-point correlation, and by less than 3\% for $k<0.5$ $\kmpc$. In Fourier space, the two power spectra agree everywhere within 5\%. The quadrupole moment of the two-point correlation function and the power spectrum is also very close to $0$ at $s$ = 10 $\smpc$ or for all $k$ modes, indicating that the spatial anisotropy caused by redshift-space distortions has been well-removed.

Our method 
does not require interpolating the velocity field at galaxies and relies less on physical quantities, which is suitable for being extended from simulations to observations, given that the properties of our mocked galaxies agree well with the observational data statistically.  Our method has potentially wide applications in cosmology, such as obtaining the cosmic velocity field from redshift space and eliminating the nonlinear effect of redshift distortion in the BAO reconstruction.

{In this study, we introduce a straightforward physical model to forecast the velocities of galaxies based on their surrounding environment. Indeed, this model shows promise for application to real observational data to determine the line-of-sight velocity of any observed galaxy. Additional testing is necessary and would be valuable for future exploration.}

For the trained \ourModel to be applicable in real observations, it has to be adaptable to different cosmologies and with different types of galaxies. {Our test with the galaxy catalogues populated by a semi-analytic galaxy formation model, indicates that our networks are not dependent on the specific galaxy formation model employed.} Our tests with the high redshift and high bias mock galaxy samples suggest that on large scales ($s>15$ $\smpc$), it is successful in handling the dependence on the growth-rate parameter -- which is the key parameter connecting the density and velocity field. However, it fails on smaller scales. We think that this is likely because of the strong non-linear effect, the Finger-of-God, has a different dependence on the growth rate parameter than in the linear regime, and our current model is not able to capture it. In a second paper, we will attempt to model the small-scale virial velocity in galaxy clusters and provide a solution to the small-scale \fog effect. Thus, the RSD effect will be well modeled on both large and small scales, and we hope to provide an efficient and accurate way to eliminate the uncertainties on the line-of-sight distance caused by the RSD effect in real galaxy surveys.

\section*{Acknowledgements}
{We also thank the anonymous reviewer for his/her many suggestions for improving this paper.} This work is supported by the National Key R\&D Program of China ( 2022YFA1602901), the NSFC grant (Nos. 11988101, 11873051, 12125302, 11903043), CAS Project for Young Scientists in Basic Research Grant (No. YSBR-062), and the K.C. Wong Education Foundation. YC acknowledges the support of
the UK Royal Society through a University Research Fellowship.  For the purpose of open access, the author has applied a Creative Commons Attribution (CC BY) licence to any Author Accepted Manuscript version arising from this submission.

\section*{Data Availability}
The data underlying this article will be shared on reasonable request to the corresponding author.
 



\bibliographystyle{mnras}
\bibliography{example} 



\bsp	
\label{lastpage}
\end{document}